\documentclass[twocolumn,superscriptaddress,amsmath,amssymb]{revtex4}
\usepackage{amsmath}
\usepackage{amssymb}
\usepackage{graphicx}
\usepackage{dcolumn}
\usepackage{bm}
\usepackage{hyperref}
\usepackage{mathrsfs} 
\usepackage{gensymb} 
\usepackage[absolute,overlay]{textpos}

\begin{document}

\title{Orbital dichotomy of Fermi liquid properties in Sr\textsubscript{2}RuO\textsubscript{4} revealed by Raman spectroscopy}

\author{Jean-Côme Philippe}
\email{jeancome.philippe@u-paris.fr}
\affiliation{Université de Paris, Matériaux et Phénomènes Quantiques, UMR CNRS 7162, Bâtiment Condorcet, 75205 Paris Cedex 13, France}
\author{Benoît Baptiste}
\affiliation{Sorbonne Université, Institut de Minéralogie, de Physique des Matériaux et de Cosmochimie, UMR CNRS 7590, IMPMC, 4 place Jussieu, 75005 Paris, France}
\author{Chanchal Sow}   
\author{Yoshiteru Maeno}
\affiliation{Department of Physics, Kyoto University, Kyoto, Japan}
\author{Anne Forget}
\author{Dorothée Colson}
\affiliation{Service de Physique de l'Etat Condensé, DSM/DRECAM/SPEC, CEA Saclay, Gif-sur-Yvette, 91191, France}
\author{Maximilien Cazayous}
\author{Alain Sacuto}
\affiliation{Université de Paris, Matériaux et Phénomènes Quantiques, UMR CNRS 7162, Bâtiment Condorcet, 75205 Paris Cedex 13, France}
\author{Yann Gallais}
\email{yann.gallais@u-paris.fr}
\affiliation{Université de Paris, Matériaux et Phénomènes Quantiques, UMR CNRS 7162, Bâtiment Condorcet, 75205 Paris Cedex 13, France}

\date{\today}

\begin{abstract}
We report a polarization-resolved Raman spectroscopy study of the orbital dependence of the quasiparticles properties in the prototypical multi-band Fermi liquid Sr\textsubscript{2}RuO\textsubscript{4}. We show that the quasiparticle scattering rate displays $\omega^{2}$ dependence as expected for a Fermi liquid. Besides, we observe a clear polarization-dependence in the energy and temperature dependence of the quasiparticle scattering rate and mass,  with the $d_{xz/yz}$ orbital derived quasiparticles showing significantly more robust Fermi liquid properties than the $d_{xy}$ orbital derived ones. The observed orbital dichotomy of the quasiparticles is consistent with the picture of Sr\textsubscript{2}RuO\textsubscript{4} as a Hund's metal. Our study establishes Raman scattering as a powerful probe of Fermi liquid properties in correlated metals.

\end{abstract}

\maketitle

\section*{\label{sec:Introduction}Introduction}

More than twenty-five years after the beginning of intensive research on Sr\textsubscript{2}RuO\textsubscript{4}, following the discovery of its superconductivity~\cite{maeno_superconductivity_1994}, this material still exhibits a two-face status. On the one hand, despite numerous theoretical and experimental results, its superconducting state developing under 1.5~K remains a puzzle, with no definitive outcome about its parity and the gap structure~\cite{mackenzie_even_2017,pustogow_constraints_2019}. On the other hand, its normal state up to $T_{FL} \approx$~25~K is now described as the prototype of a quasi-two-dimensional Fermi liquid (FL)~\cite{bergemann_quasi-two-dimensional_2003}, so that "it is an established \textit{unconventional} superconductor, with a \textit{conventional} low-temperature normal state"~\cite{bergemann_quasi-two-dimensional_2003}.
\par
The FL behavior is experimentally well established through the analysis of electron transport and optical conductivity measurements which display hallmarks of electron-electron interactions for a correlated FL state: the $\omega^2$ and $T^2$ dependencies of scattering rate reflecting the low energy phase space constraints on electron-electron collisions. $T^{2}$ dependence of the in-plane DC resistivity is observed below $T_{FL} \sim 25$~K~\cite{maeno_normal-state_1996}. The corresponding $\omega^2$ dependence of the scattering rate has been somewhat more elusive experimentally. Early Angle Resolved Photo-Emission Spectroscopy (ARPES) data indicated $\omega^2$ dependence of the quasiparticle scattering rate, but the reported inverse lifetimes exceed the quasiparticle energies even at low energy, indicating they are likely not representative of the intrinsic bulk properties \cite{ingle_quantitative_2005, kidd_orbital_2005}. Still, optical conductivity data show that in Sr\textsubscript{2}RuO\textsubscript{4} the Gurzhi scaling law relating the $T^{2}$ and $\omega^{2}$ prefactors of the scattering rate of a FL ~\cite{gurzhi_mutual_1959} is experimentally verified~\cite{stricker_optical_2014,maslov_optical_2017}. Large effective masses $m^{*}$ ranging from 3 to 5 times the band mass were measured via specific heat~\cite{maeno_two-dimensional_1997}, quantum oscillations ~\cite{mackenzie_fermi_1998, bergemann_quasi-two-dimensional_2003}, ARPES \cite{tamai_high-resolution_2019} and optical reflectivity measurements ~\cite{katsufuji_in-plane_1996, lee_orbital-selective_2006, stricker_optical_2014}, which indicate significant electron correlations. Rather than the proximity to a Mott insulator, the Hund's rule coupling has been identified as the origin of heavy quasiparticle mass through Dynamical Mean Field Theory (DMFT) calculations~\cite{mravlje_coherence-incoherence_2011,georges_strong_2013}, labelling Sr\textsubscript{2}RuO\textsubscript{4} as a Hund's metal.
\par
The Fermi surface of this material is known in-depth from quantum oscillations~\cite{mackenzie_quantum_1996, bergemann_detailed_2000} and ARPES~\cite{damascelli_fermi_2000, tamai_high-resolution_2019}. It presents three quasi-cylindrical sheets $\alpha$, $\beta$ and $\gamma$: $\alpha$ and $\beta$ come mainly from the hybridization of the one-dimensional $d_{xz}$ and $d_{yz}$ orbitals, whereas $\gamma$ corresponds primarily to the two-dimensionnal $d_{xy}$ orbital (Fig. \ref{fig:Raman Vertex}). The multi-orbital nature of Sr\textsubscript{2}RuO\textsubscript{4} induces some complications in the understanding of the experimental results since transport and optical conductivity measurements average the current response of the different Fermi sheets. 
Insight on the orbital dependent FL properties, a key feature of a Hund's metal, has been mostly limited to the static mass-enhancement factor which has been studied by quantum oscillations and ARPES measurements ~\cite{bergemann_detailed_2000,tamai_high-resolution_2019}. Both indicate that the $d_{xy}$ orbital derived quasiparticles are more correlated than the $d_{xz}$ and $d_{yz}$ ones, with larger static mass-enhancement, in agreement with DMFT calculations ~\cite{mravlje_coherence-incoherence_2011, kugler_strongly_2020}. Note that a similar orbital differentiation is also found in the magnetic sector, with possible implications for the still unsolved superconducting pairing mechanism of Sr\textsubscript{2}RuO\textsubscript{4} \cite{mazin_competitions_1999,romer_knight_2019}: whereas antiferromagnetic fluctuations arise from the nested $d_{xz}$ and $d_{yz}$ derived bands, sub-leading ferromagnetic fluctuations are attributed to the $d_{xy}$ derived band \cite{sidis_evidence_1999, steffens_spin_2019}. The coupling between the magnetic fluctuations is expected to lead to distinctive features in the orbital, wave-vector and energy dependence of the quasiparticle properties. A potential benchmark to test the respective influence of Hund's rule coupling physics and low energy magnetic fluctuations can thus be reached by probing the orbital dependence of the quasiparticle dynamics in the FL state. 
\par

In this paper, we use polarization-resolved Raman spectroscopy to access the low energy quasiparticles dynamics of Sr\textsubscript{2}RuO\textsubscript{4} in an orbital-resolved way. We show that the quasiparticle scattering rate displays $\omega^{2}$ dependence as expected for a FL, thus mirroring the $T^{2}$ dependence observed in transport measurements. Besides, we observe a clear polarization-dependence in the energy and temperature dependence of the quasiparticle scattering rate and mass. We assign it to orbitally dependent FL properties with the $d_{xy}$ derived quasiparticles being significantly more correlated than the $d_{yz}$/$d_{xz}$ derived ones. The observed dichotomy is consistent with the picture of Sr\textsubscript{2}RuO\textsubscript{4} as a Hund's metal with significant orbital differentiation among quasiparticles. Our study establishes Raman scattering as a powerful probe of FL properties in multiband correlated metals.

\section*{\label{sec:Experiment}Experiments}

Raman experiments have been carried out using a triple grating JY-T64000 spectrometer in subtractive configuration with 1800 grooves/mm gratings. For measurements above 75~meV (discussed in the supplementary information (SI)~\cite{SM}), a single-stage configuration was used with 600 grooves/mm grating. When using the 1800 grooves/mm configuration, measurements could be performed down to 1~meV, with a resolution of 0.2~meV. The spectrometer is equipped with a nitrogen cooled CCD detector. We used the 532~nm excitation line from a diode pump solid state laser. Additional spectra displayed in SI~\cite{SM} were obtained by using 488~nm and 660~nm wavelength lasers. Measurements between 3 and 200~K have been performed using a closed-cycle optical cryostat. Comparing temperature-dependent and laser-power-dependent spectra, we estimate laser heating to be $\Delta T \approx 4$~K \cite{SM}. All the raw spectra have been corrected for the Bose factor and the instrumental spectral response. They are thus proportional to the imaginary part of the Raman response function $\chi^{\prime\prime}(\omega,T)$.
\par
As Sr\textsubscript{2}RuO\textsubscript{4} belongs to the $D_{4h}$ point group, the Raman-accessible symmetries are $A_{1g}$, $E_{g}$, $B_{1g}$ and $B_{2g}$. For $B_{2g}$ and $B_{1g}$ symmetries measurements, the direction of incoming and outgoing electric fields are contained in the (\textit{ab}) plane, with crossed polarizations along and at 45$\degree$ from the Ru–O bond directions, respectively. For $A_{1g}$ symmetry measurements presented here, the polarizations are parallel along the (\textit{c}) axis (so we probe the out-of-plane component of the $A_{1g}$ channel). For $E_{g}$ symmetry measurements, the direction of incoming electric field is along the (\textit{c}) axis, the outgoing one is in the (\textit{ab}) plane. 

The single crystal of Sr\textsubscript{2}RuO\textsubscript{4} used in our experiment was grown by the floating zone technique as described elsewhere~\cite{bobowski_improved_2019}. Two different samples cut from the same single crystal were studied. One with an (\textit{ab}) surface was used to access $B_{2g}$ and $B_{1g}$ symmetries, while the other with an (\textit{ac}) surface was used to access $A_{1g} (c)$ and $E_{g}$ symmetries. The second sample was wire-sawed from the crystal and so required surface polishing before performing spectroscopy. The crystallographic axes were determined via single-crystal X-ray diffraction prior to the Raman measurements.

\section*{Orbital dependence of Raman scattering symmetries in Sr\textsubscript{2}RuO\textsubscript{4}}
\begin{figure}
\includegraphics[width=0.485\textwidth]{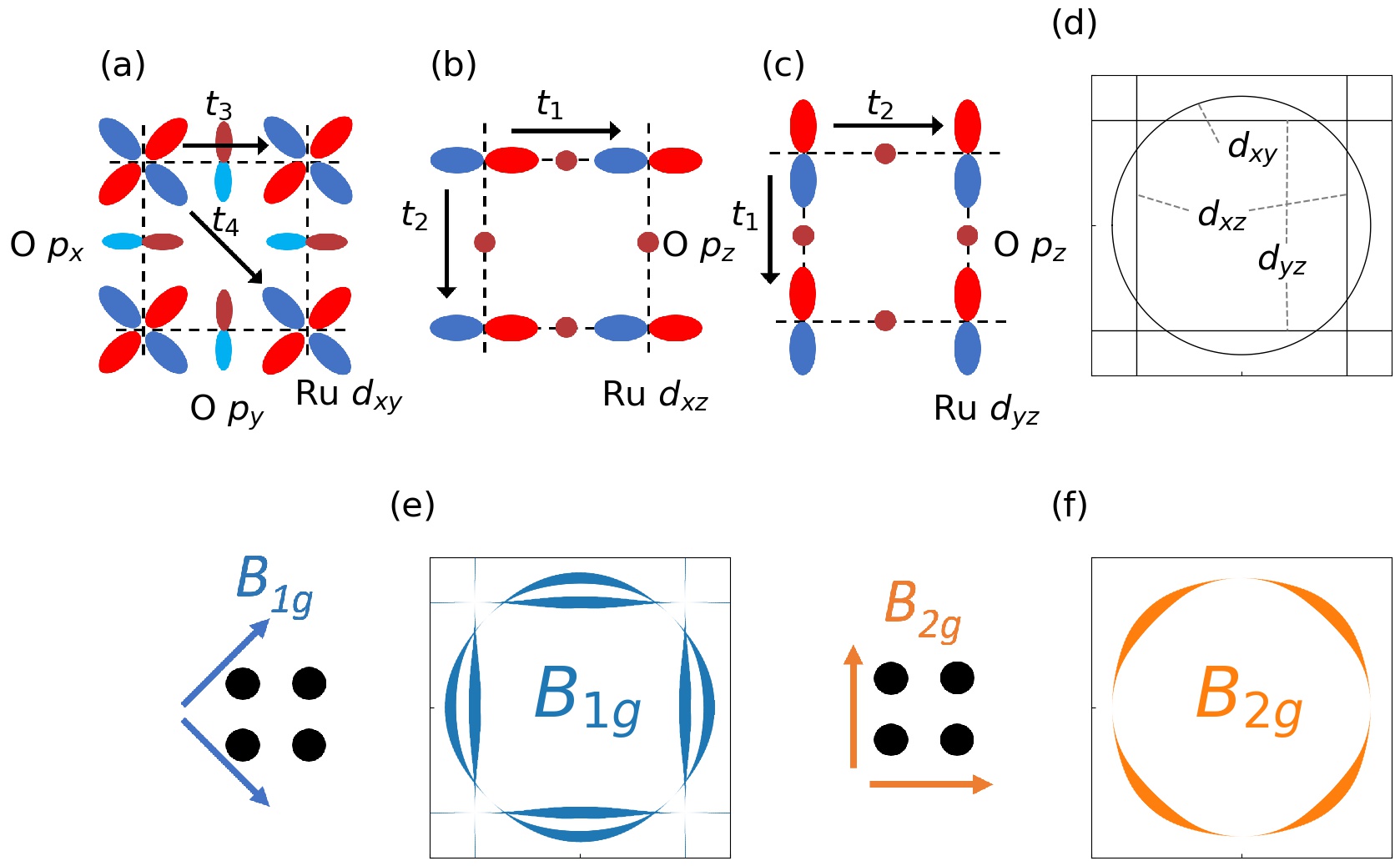}
\caption{\label{fig:Raman Vertex} (a,b,c)~Schematic representation of the two-dimensional intra-orbital hopping integrals of the Ru $d$ orbitals, $t_{i}$. (d)~Schematic representation of the Fermi surface of Sr\textsubscript{2}RuO\textsubscript{4} in the first Brillouin zone. (e,f)~ Sketch of the wave-vector dependence on the Fermi surface of the squared Raman vertex for intra-orbital transition terms in  $B_{1g}$ and $B_{2g}$ channels. The local linewidth denotes the strength of the vertex. Note the nodes in the Raman vertex value along the diagonals or along the principal axis directions for $B_{1g}$ and $B_{2g}$ channels respectively. We took the values of $t_{i}$ as $\{t_{1},t_{2},t_{3},t_{4}\} = \{88,9,80,40\}$~meV~\cite{cobo_anisotropic_2016}. The sketches on the left of each vertex depict for each symmetry the polarizations of the incident and scattered beams with respect to the in-plane Ru square lattice.}
\end{figure}

The main conclusions of this work are based on the orbital selectivity of the $B_{1g}$ and $B_{2g}$ symmetries which we discuss first (more details can be found in SI~\cite{SM}).
Within the effective mass approximation the amplitude of the Raman response, or Raman vertex, $\gamma^{\mu}(\mathbf{k})$ is symmetry dependent and given by the second derivatives of the band dispersion \cite{devereaux_inelastic_2007}. In orbital space ($a=xz/yz/xy$), the $B_{1g}$ and $B_{2g}$ Raman vertices corresponding to intra-orbital excitations are given by $\gamma_{a}^{B_{1g}}(\mathbf{k})=\frac{1}{2}(\frac{\partial^2\epsilon_{a}}{\partial k_x^2}-\frac{\partial^2\epsilon_{a}}{\partial k_y^2})$ and $\gamma_{a}^{B_{2g}}(\mathbf{k})=\frac{\partial^2\epsilon_{a}}{\partial k_x \partial k_y}$. Using the hopping terms coming from the tight-binding model applicable to Sr\textsubscript{2}RuO\textsubscript{4} ~\cite{cobo_anisotropic_2016, romer_knight_2019}, with hopping integrals $t_{i}$, we obtain $\gamma_{xz/yz}^{B_{1g}}(\mathbf{k})=t_{1/2}\cos(k_{x})-t_{2/1}\cos(k_{y})$, $\gamma_{xy}^{B_{1g}}(\mathbf{k})=t_{3}(\cos(k_{x})-\cos(k_{y}))$, $\gamma_{xz/yz}^{B_{2g}}(\mathbf{k})=0$ and $\gamma_{xy}^{B_{2g}}(\mathbf{k})=4t_{4}\sin(k_{x})\sin(k_{y})$.
Here the dominant hopping terms are $t_1$ the nearest neighbour Ru-Ru hopping between $xz$ and $yz$ orbitals along $x$ and $y$ respectively, and the nearest $t_3$ and next nearest $t_4$ neighbour Ru-Ru hopping between $xy$ orbitals. The intra-orbital hoppings and the resulting Fermi surface are sketched in Fig.~\ref{fig:Raman Vertex}-(a-d). While the $k$ dependent form factors are imposed by symmetry, the vanishing contribution of the $d_{xz}$/$d_{yz}$ orbital in $B_{2g}$ stems from the absence of next-nearest neighbour hopping between $xz$/$yz$ orbitals. The orbital and $k$ dependence of the vertices are sketched in Fig.~\ref{fig:Raman Vertex}-(e,f). In this simple picture, the $B_{1g}$ channel probes intra-orbital excitations for both $d_{xz/yz}$ and $d_{xy}$ orbitals with a priori similar weights ($t_1\sim t_3$), providing limited orbital resolution. By contrast the $B_{2g}$ channel only probes $d_{xy}$ intra-orbital excitations, giving a unique access to the properties of the quasiparticles arising from this orbital. We note that inter-orbital excitations between $d_{xz}$ and $d_{yz}$ will also contribute to the $B_{2g}$ channel, but with a weight about two orders of magnitude smaller than intra-orbital excitations due to the smallness of the inter-orbital hopping integrals \cite{SM}.

\section*{\label{sec:FL}Raman response of a Fermi liquid}

\begin{figure}
\includegraphics[width=0.495\textwidth]{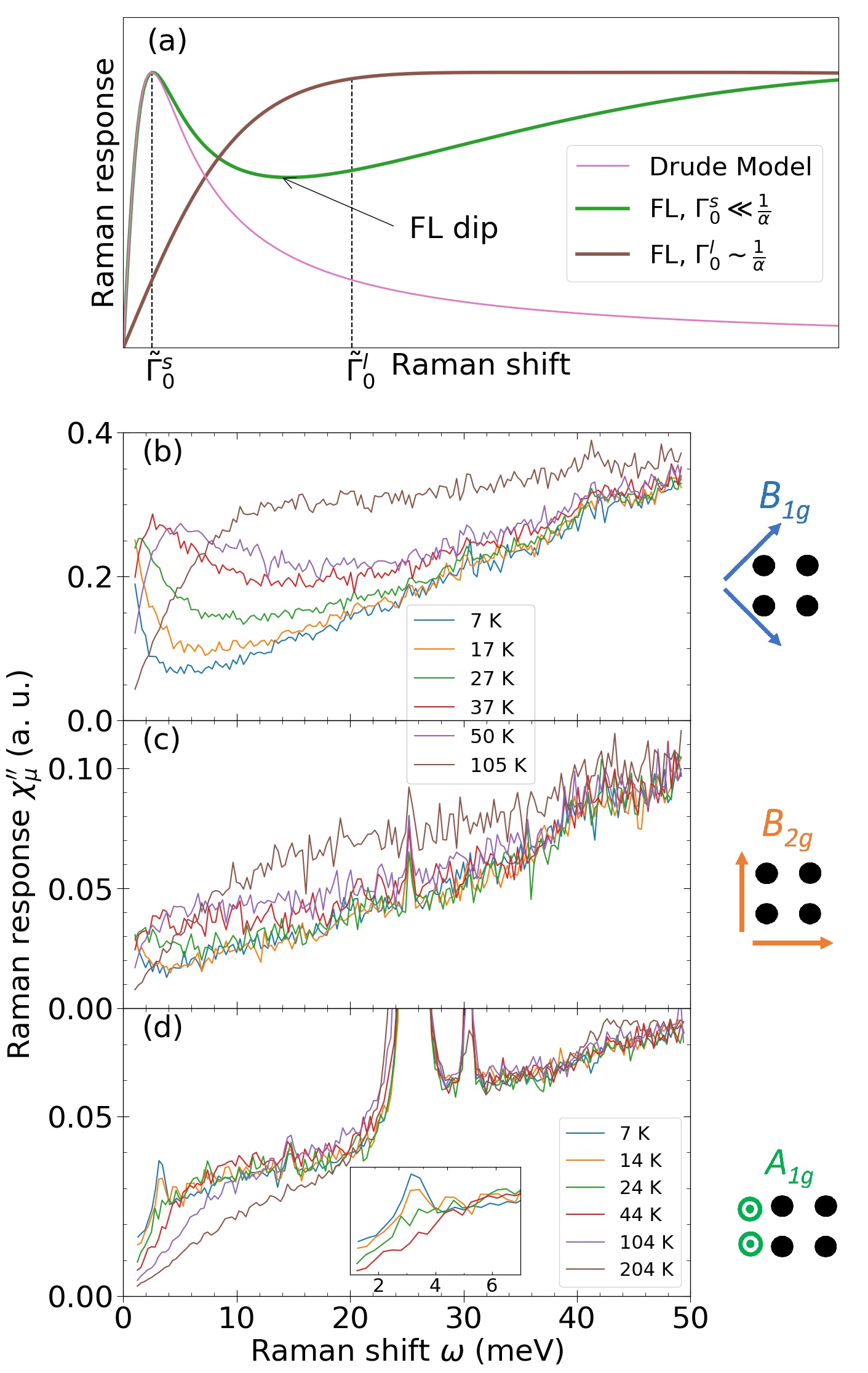}
\caption{\label{fig:Raman responses Theory and Expe} (a) Theoretical Raman responses expected for a metal in the Drude model and for a FL in the extended Drude model. We took the small and the large $\tilde{\Gamma}_{0}$ as $\tilde{\Gamma}^{l}_{0} = 8\tilde{\Gamma}^{s}_{0}$. For the Drude model $\Gamma_{0}$ is taken equal to $\tilde{\Gamma}^{s}_{0}$. (b,c,d) Experimental Raman responses obtained in the $B_{1g}$, $B_{2g}$ and $A_{1g}$ symmetry channels at selected temperatures. The sketches on the right depict for each symmetry the polarizations of the incident and scattered beams with respect to the in-plane Ru square lattice. In the (d) panel, the inset shows the $A_{1g}$ response below 7~meV.}
\end{figure}

Having discussed the orbital selectivity of Raman scattering, we now discuss the theoretically expected spectral lineshape of the Raman spectra of a FL which to our knowledge has not been discussed previously. We drop the temperature dependence for this discussion. In a simple Drude model the electronic Raman response reads \cite{ipatova_electron_nodate, zawadowski_theory_1990}:

\begin{equation}
\chi^{\prime\prime}_{\mu}(\omega) = \chi_{\mu}^{0}\frac{\omega\Gamma_{\mu}}{\omega^{2} + \Gamma_{\mu}^{2}}
\label{reponse_raman_DM}
\end{equation}
with the index $\mu$ standing for the Raman symmetry channel, $\Gamma_{\mu}$ the symmetry resolved electronic scattering rate and $\chi_{\mu}^{0}$ the static susceptibility. In the Drude model, $\Gamma_{\mu}$ shows no energy dependence and can be assigned to static impurities. The Raman response displays a Drude peak at $\omega = \Gamma_{\mu}$ and then decreases with energy vanishing when $\omega>>\Gamma_{\mu}$ (Fig. \ref{fig:Raman responses Theory and Expe}-(a)). For a FL with energy-dependent quasiparticle scattering rate $\Gamma_{\mu}(\omega)$ and mass-enhancement factor $1+\lambda_{\mu}$, in close analogy with the current response of optical conductivity, the electronic Raman response can be modelled by the extended Drude model (EDM) (see also SI~\cite{SM} for a discussion of the validity of the EDM description for the Raman response):

\begin{equation}
\chi^{\prime\prime}_{\mu}(\omega) = \tilde\chi_{\mu}^0\frac{\omega\tilde{\Gamma}_{\mu}(\omega)}{\omega^{2} + \tilde{\Gamma}_{\mu}(\omega)^{2}}
\label{reponse_raman_EDM}
\end{equation}
where $\tilde{\chi}^0_{\mu}=\frac{\chi^0_{\mu}}{1+\lambda_{\mu}}$ and $\tilde{\Gamma}_{\mu}(\omega)=\frac{\Gamma_{\mu}(\omega)}{1+\lambda_{\mu}}$. At low energy and below $T_{FL}$ we expect $\Gamma_{\mu}(\omega)=\Gamma_{\mu,0} + \alpha_{\mu}\omega^{2}$ for the scattering and a constant mass-enhancement 1+$\lambda_{\mu}$. The main difference in the Raman response between Drude model and EDM is the non-monotonic lineshape of the response beyond the Drude peak, leading to a characteristic FL "dip" in the Raman spectra separating the Drude regime at lower energy, and the thermal regime at higher energy in close analogy with the FL "foot" of optical conductivity~\cite{berthod_non-drude_2013}. Increasing the static $\Gamma_{\mu,0}$ by increasing temperature and/or disorder, the Drude peak hardens, causing the FL dip to resorb and making the Raman response more Drude-like. The disappearance of the FL dip occurs when $\Gamma_{\mu,0}\sim \frac{1}{\alpha}$. The presence of the FL dip in the Raman spectrum is a clear fingerprint of FL behavior, which to our knowledge has not been reported up to now. Instead, the flat Raman continua observed in correlated metals like cuprates~\cite{opel_carrier_2000}, and iridates~\cite{sen_strange_2020} have been interpreted as signaling non-FL "marginal" like behavior, with a quasiparticle scattering rate following linear rather than quadratic energy dependence. From this perspective Sr\textsubscript{2}RuO\textsubscript{4} provides an interesting model system to establish canonical FL behavior in the Raman scattering spectrum.

\section*{\label{sec:Results}Results}

The spectra obtained in $B_{1g}$, $B_{2g}$ and $A_{1g}$ symmetries are depicted in Fig. ~\ref{fig:Raman responses Theory and Expe}-(b,c,d) at selected temperatures (more temperatures for $B_{1g}$ and $B_{2g}$ can be found in SI~\cite{SM}). As discussed previously, ($ab$) plane-polarized $B_{1g}$ and $B_{2g}$ symmetries probe in-plane quasiparticle excitations arising from $d_{xz}$, $d_{yz}$ and $d_{xy}$ orbitals. On the other hand the $c$-axis polarized $A_{1g}$ symmetry probes out-of-plane quasiparticle excitations related to the weak interlayer hopping process. The spectra obtained in mixed $E_{g}$ symmetry are shown in the SI~\cite{SM}. In $B_{1g}$ and $B_{2g}$ symmetries, the spectra are dominated by the electronic continuum and are essentially free from any phononic contribution, as expected for the I4/mmm space group of Sr\textsubscript{2}RuO\textsubscript{4}. Besides the continuum, the $A_{1g}$ spectra also display narrow peaks assigned to the $A_{1g}$ optical phonon modes coming from $c$-axis motion of Sr atoms, and a leakage from the $E_{g}$ phonon coming from $ab$ plane motion of Sr atoms respectively ~\cite{udagawa_phonon_1996, sakita_anisotropic_2001, iliev_comparative_2005}
\par
In all symmetries  the signal remains unchanged up to at least 200~K above around 50~meV, but we observe a significant temperature dependence of the low energy response. In $B_{1g}$ symmetry the shape of the Raman spectrum below 100~K is consistent with the response expected for a FL described above. It contains both a low energy Drude peak and a FL dip at higher energy. The Drude peak hardens and broadens and the FL dip resorbs as temperature increases, in qualitative agreement with a static relaxation rate $\Gamma_{B_{1g},0}$ increasing with temperature. In $B_{2g}$ the temperature evolution is qualitatively similar, but with key differences: the Drude peak is strongly reduced with respect to the continuum at higher energy. A weaker FL dip can still be resolved but below 50~K instead of 85~K at least in the $B_{1g}$ channel. Below 10~K and in both symmetries, the Drude peak maximum shifts well below our low energy cutoff at 1~meV, indicating an extremely low residual scattering rate due to disorder. 
\par
The $A_{1g}$ response is qualitatively different from the in-plane symmetries, with an almost flat continuum in the whole spectral range. Above 20~K no clear FL dip is resolved and a much broader Drude peak is observed, indicating a significantly larger static relaxation rate in this out-of-plane channel, consistent with transport measurements \cite{hussey_normal-state_1998}. Below 20~K, a narrow peak develops around 3.5~meV. This peak is too sharp and symmetric to be assigned to a Drude-like response, or an interband transition. Instead, we tentatively assign a collective excitation, possibly a $c$-axis plasmon, which would correspond to the onset of coherent $c$-axis transport observed in resistivity measurements below $\sim$ $T_{FL}$ ~\cite{hussey_normal-state_1998}. We note that recent momentum-resolved electron energy loss spectroscopy (M-EELS) measurements in Sr\textsubscript{2}RuO\textsubscript{4} report a dispersive collective mode of electronic origin. The $q=0$ intercept of the M-EELS mode falls below 10~meV, and could therefore correspond to the same excitation as the one observed in our $c$-axis $A_{1g}$ Raman spectrum~\cite{husain_coexisting_2020}. 
\par
In the following we will focus on the FL analysis of the in-plane quasiparticle dynamics obtained from the $B_{1g}$ and $B_{2g}$ spectra.

\section*{\label{sec:Analysis}Analysis of the symmetry-resolved Fermi liquid properties}

From the $B_{1g}$ and $B_{2g}$ Raman spectra we can extract the symmetry resolved relaxation rate $\Gamma_{\mu}$ and mass-enhancement factor $1 + \lambda_{\mu}$ using the memory-function approach~\cite{gotze_homogeneous_1972, opel_carrier_2000}. This method, widely used for the analysis of the optical conductivity spectrum of correlated electron systems, was first introduced by Opel et al. \cite{opel_carrier_2000} to analyze the normal state Raman response of high-T$_c$ cuprates. It is described in details in the SI~\cite{SM}. For a FL, the relaxation rate is expected to follow a quadratic behavior both in energy and temperature, below some crossover temperature $T_{FL}$ and energy $\omega_{FL}$.

\begin{equation}
\Gamma_{\mu}(\omega,T)=\Gamma_{\mu,00} + \alpha_{\mu} \omega^{2} + \beta_{\mu}(k_{B} T)^{2}
\end{equation}

\begin{figure}
\includegraphics[width=0.485\textwidth]{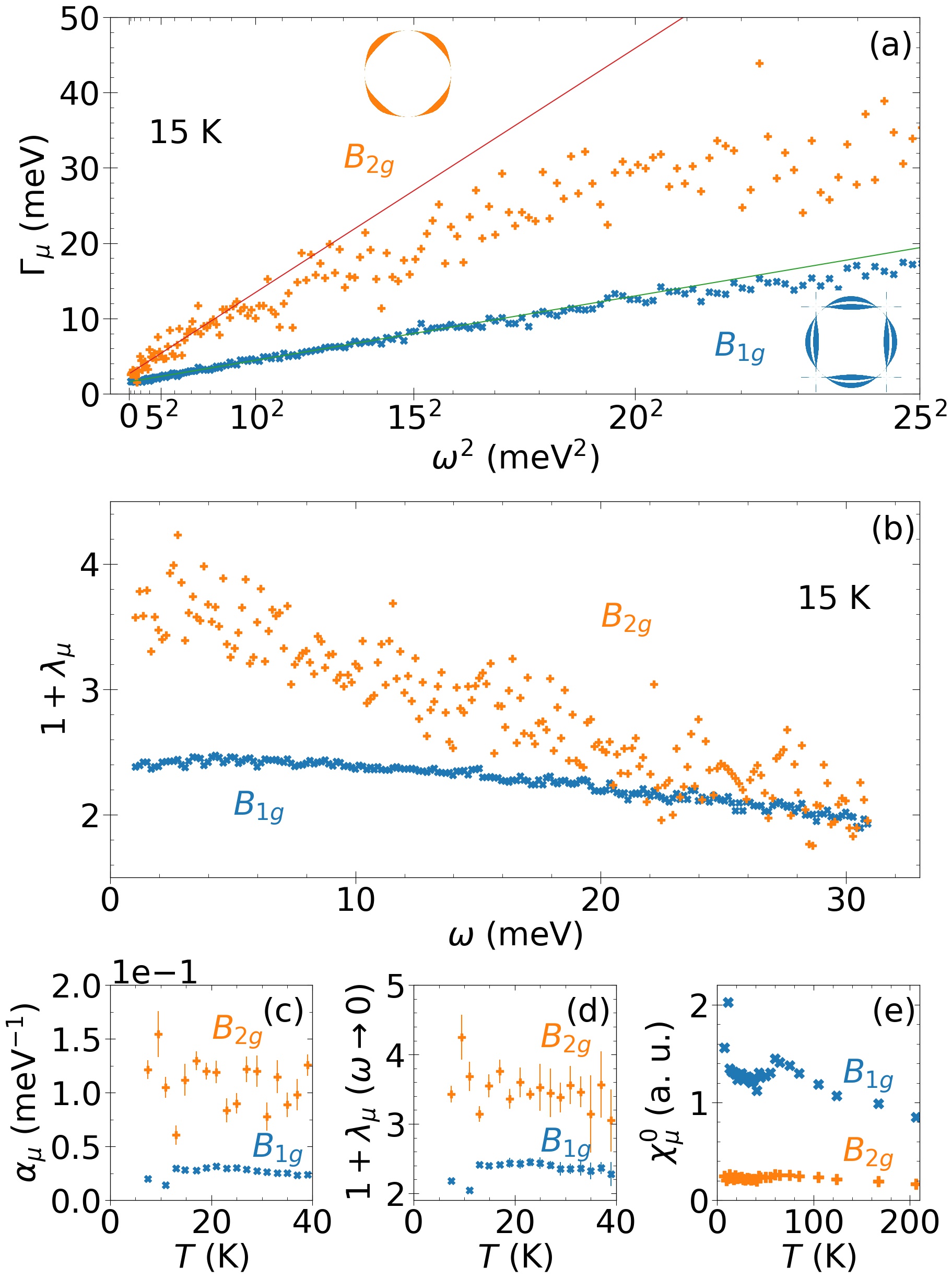}
\caption{\label{fig:Results in frequency}(a)~Energy dependence of the relaxation rate $\Gamma_{\mu}(\omega)$ at 15~K for $B_{1g}$ and $B_{2g}$. The solid lines are quadratic fits of the experimental data at low energy. The sketches remind the shape of the Raman vertices (see Fig.~\ref{fig:Raman Vertex}). (b)~Low energy dependence of the mass-enhancement factor $1 + \lambda_{\mu}$ at 15~K in $B_{1g}$ and $B_{2g}$. (c)~Temperature dependence of the $\omega^{2}$ prefactor of $\Gamma_{\mu}(\omega)$, obtained through the fits. The error bars are computed as statistical uncertainty in the quadratic fits. (d)~Temperature dependence of the extrapolated zero-energy mass-enhancement. (e)~Temperature dependence of the Raman static susceptibility.}
\end{figure}

First we consider the energy dependence of $\Gamma_{\mu}$ and $1 + \lambda_{\mu}$ at 15~K, deep in the FL transport regime  (Fig.~\ref{fig:Results in frequency}). As shown in Fig. \ref{fig:Results in frequency}-(a,b), there is a significant symmetry dependence of both quantities with  $\Gamma_{B_{1g}} < \Gamma_{B_{2g}}$ and $1 + \lambda_{B_{1g}} < 1 + \lambda_{B_{2g}}$ over the whole relevant energy range. Also, the scattering rate follows $\omega^2$ behavior over a significantly broader energy range in $B_{1g}$ symmetry compared to $B_{2g}$ symmetry. We estimated the range of $\omega_{FL,\mu}$ by performing quadratic fits of low energy $\Gamma_{\mu}(\omega)$ and tracking the energy at which a departure from quadratic behavior is resolved (see SI~\cite{SM} for further details). Considering all probed temperatures below 40~K, we obtain $\omega_{FL,B_{1g}} \sim 15-20$~meV and $\omega_{FL,B_{2g}} \sim 8-12$~meV, with a clear quantitative dichotomy between the two channels.

From the quadratic fits we obtain the $\omega^2$ FL coefficient $\alpha_{\mu}$ in both symmetries. As shown in Fig.~\ref{fig:Results in frequency}-(c) $\alpha_{\mu}$ shows a weak temperature dependence in both symmetries, and despite larger uncertainties in $B_{2g}$ symmetry, $\alpha_{B_{2g}} > \alpha_{B_{1g}}$ holds in the entire temperature range. The greater dispersion below 12~K is linked with the fact that in this temperature range the Drude peak is below the low energy experimental Raman cut-off. This causes difficulties in accurately extrapolating the experimental data to zero energy, and so a greater uncertainty in all the extracted quantities (see SI for further details on the spectra low energy extrapolation~\cite{SM}). To estimate quantitative values of $\alpha_{\mu}$ deep in the FL state, we consider values between 12 and 25~K. Overall, taking the average of the values in this temperature range we obtain $\alpha_{B_{2g}} \approx 3\alpha_{B_{1g}}$.
\par
The mass-enhancement factor decreases with energy in both symmetries and stays larger than unity at high energy. It is flatter at low energy in $B_{1g}$ symmetry as expected for a FL where it should be constant below $\omega_{FL}$ \cite{berthod_non-drude_2013}, and in agreement with the more robust FL behavior observed in the scattering rate in the same symmetry. The extrapolated values at low energy $1 + \lambda_{\mu}(\omega \rightarrow 0)$ are weakly temperature dependant below 40~K but display a significant symmetry dependence, with $1 + \lambda_{0}^{B_{2g}} \approx 1.5 (1 + \lambda_{0}^{B_{1g}})$, and values ranging from 2.5 to 4 (Fig.~\ref{fig:Results in frequency}-(d)). 

Finally the symmetry-resolved static susceptibility $\chi^{0}_{\mu}$ (Fig.~\ref{fig:Results in frequency}-(e)) displays a mild dependence on temperature with a gradual increase in both symmetry channels upon cooling from 200~K to 60~K, followed by an essentially flat temperature dependence below 60~K. This indicates the absence of any significant symmetry breaking nematic-like instability in the $B_{1g}$ or $B_{2g}$ channels. 

\par
Focusing now on the temperature dependence of the scattering rate, we define the static relaxation rate $\Gamma_{\mu,0}(T)$ as the extrapolated relaxation rate at zero energy obtained through the quadratic energy fits. Its temperature dependence is depicted in figure~\ref{fig:Gamma temperature and gurzhi}-(a). We fit the $\Gamma_{\mu,0}(T)$ data below 27~K by a quadratic behavior and we estimate $T_{FL,\mu}$ as the range above which the data visually deviate from the fit. We define $\Gamma_{\mu,00}$ as the zero temperature value of the fitted $\Gamma_{\mu,0}(T)$. We have significantly less experimental points in temperature than in energy, so it is more difficult to estimate the range of $T_{FL,\mu}$. We estimate $T_{FL,B_{1g},B_{2g}} \approx 20-30$~K, and we can extract the symmetry resolved $T^2$ prefactor $\beta_{\mu}$. We cannot conclude about a dichotomy between $T_{FL,B_{1g}}$ and $T_{FL,B_{2g}}$, but, despite larger uncertainties compared to $\omega^2$ coefficient, it is clear that $\beta_{B_{2g}} > \beta_{B_{1g}}$.

The Table~\ref{table:tableau_parametres_FL_anisotropie} summarizes the FL parameters in both symmetry channels. The different values of $\alpha_{\mu}$, $\beta_{\mu}$, $\omega_{FL,\mu}$ and $1 + \lambda_{\mu}(\omega \rightarrow 0)$ suggest that the FL state is more robust in the $B_{1g}$ than in the $B_{2g}$ channel: the quasiparticles probed via $B_{2g}$ are more correlated than the ones probed via $B_{1g}$. Given the orbital dependence of each symmetry channel discussed previously, we can conclude that the electrons originating from the $d_{xy}$ orbital are significantly more correlated than the ones derived from the $d_{xz/yz}$ orbitals.

\begin{table*}[]
\begin{center}
\begin{tabular}{ c| c | c | c | c | c | c }
      & $\omega_{FL}$ (meV) & $\alpha$ ($10^{-2}$~meV$^{-1}$) & $\beta$ ($10^{-1}$~meV$^{-1}$) & $\Gamma_{00}$ (meV) & $p$ & $1 + \lambda(\omega \rightarrow 0)$\\ \hline
		 $B_{1g}$ & $15-20$ & 3.0 $\pm$ 0.2 & 6.9 $\pm$ 0.2 & 0.45 $\pm$ 0.06 & 1.6 $\pm$ 0.1 & 2.45 $\pm$ 0.05\\
		 $B_{2g}$ & $8-12$ & 10 $\pm$ 3 & 8.9 $\pm$ 0.2 & 1.16 $\pm$ 0.05 & 0.85 $\pm$ 0.1 & 3.5 $\pm$ 0.5\\

\end{tabular}
\caption{FL parameters differentiation in $B_{1g}$ and $B_{2g}$ channels.} 
\label{table:tableau_parametres_FL_anisotropie}
\end{center}
\end{table*}

\section*{\label{sec:Discussion}Discussion}

\begin{figure}
\includegraphics[width=0.48\textwidth]{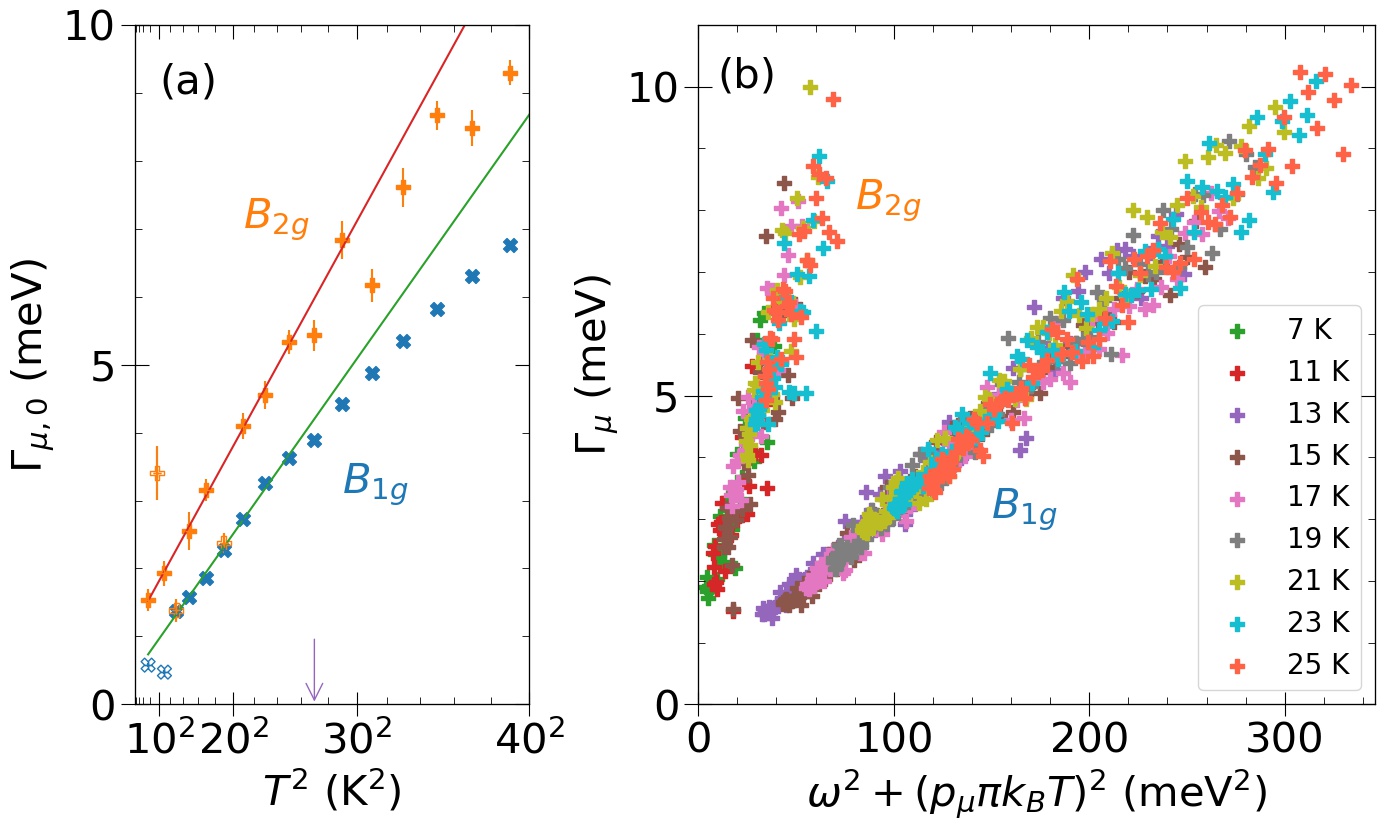}
\caption{\label{fig:Gamma temperature and gurzhi}(a)~Temperature dependence of the extrapolated static relaxation rate $\Gamma_{0,\mu}$ for $B_{1g}$ and $B_{2g}$. The solid lines are quadratic fits of the low temperature data. The arrow denotes the upper limits of the fits at 27~K. The empty symbols are excluded for the computation of the fits. (b)~Gurzhi scaling with $p_{B_{1g}} = 1.6$ and $p_{B_{2g}} = 0.85$.}
\end{figure}

The orbital differentiation in the FL properties of Sr\textsubscript{2}RuO\textsubscript{4} is consistent with previous experimental and theoretical results highlighting the more correlated status of the $\gamma$ band arising from the $d_{xy}$ orbital. 
This orbital differentiation has been mostly documented via the mass-enhancement factor through de Haas van Alphen (dHvA), and more recently ARPES measurements \cite{tamai_high-resolution_2019}. The dHvA oscillations results give the zero energy and low temperature thermodynamic cyclotron masses in each band at the Fermi level, measuring enhancement of 3.0, 3.5 and 5.5 with respect to the band masses (taken from~\cite{mcmullan_comparison_1996}) respectively for the $d_{xz/yz}$ and $d_{xy}$ derived Fermi sheets. Similar values are found by ARPES \cite{tamai_high-resolution_2019}. Our mass-enhancement values are somewhat lower, but this might be attributed to the cut-off parameter $\omega_{c}$ used to extract the Raman-derived quasiparticle properties which has a logarithmic influence on $\Gamma$ and $1 + \lambda$ making our values likely underestimated \cite{SM}.  
Still, the ratio of mass-enhancement between both Raman symmetries, 1.5, is close to the one between $d_{xy}$ and $d_{xz/yz}$ orbital derived Fermi sheets found in quantum oscillation measurements, and also in DMFT calculations \cite{mravlje_coherence-incoherence_2011}. From this, we can conclude that the $B_{1g}$ FL parameters are likely dominated by quasiparticles coming from $d_{xz/yz}$ orbitals. This is not completely intuitive since as stressed in the discussion of Raman vertices, we expect this symmetry to have roughly equal contributions from each orbitals. More realistic calculations of the Raman responses, including spin-orbit coupling and the full Fermi surface topology \cite{veenstra_spin-orbital_2014,tamai_high-resolution_2019}, are needed to further assess this point.
\par
Besides the mass enhancement, an interesting feature of our Raman results is the ability to simultaneously track the energy and temperature behaviors of the scattering rate. The link between these quantities and the mass enhancement has traditionally been discussed in terms of the Kadowaki-Woods (KW) ratio which links the mass enhancement as measured by the Sommerfeld coefficient of the electronic specific heat and the $T^2$ coefficient of resistivity measurements \cite{rice_electron-electron_1968,noauthor_universal_1986}. This ratio has been shown to be approximately constant within different class of materials like transition metals \cite{rice_electron-electron_1968} and 4f-electron heavy fermions systems \cite{noauthor_universal_1986}, indicating that it is relatively immune to the strength of electron correlations. However, the ratio is also known to depend sensitively on band structure details such as dimensionality, band-width and Fermi velocities \cite{hussey_non-generality_2005,jacko_unified_2009}. An additional source of complication in multiband systems with sizeable Hund's rule coupling is the impact of orbital differentiation on the KW ratio~\cite{cavanagh-2015}. Indeed transport measurements are dominated by high velocity bands, while specific heat is dominated by $d_{xy}$ derived $\gamma$ sheets. Insight into this orbital dependence can be obtained by computing similar ratios for both symmetries using the extracted mass-enhancement and the $\omega^2$ or $T^2$ coefficient of the scattering rate. These ratios display sizeable symmetry dependence $(\frac{\alpha}{(1+\lambda)^2})_{B_{1g}/B_{2g}}\sim5/8$~eV$^{-1}$ and $(\frac{\beta}{(1+\lambda)^2})_{B_{1g}/B_{2g}}\sim115/72$~eV$^{-1}$ pointing to an orbital dependence of such ratio within the same material. We note that a strong variation of the KW ratio was recently found in Sr\textsubscript{2}RuO\textsubscript{4} under uni-axial strain. It was attributed to the proximity of the d$_{xy}$ derived $\gamma$ sheet to a van Hove singularity in the density of state \cite{barber_resistivity_2018}. These findings confirm the strong sensitivity of such ratios to material specific band structure details such as orbital content and density of states.

\par
Another hallmark of FL behavior is the scaling between the $T^2$ and $\omega^2$ coefficients. For a FL
$\alpha_{\mu}$ and $\beta_{\mu}$ are related through the Gurzhi scaling law \cite{gurzhi_mutual_1959,maslov_first-matsubara-frequency_2012}:
\begin{equation}
\frac{\beta_{\mu}}{\alpha_{\mu}}=(p_{\mu}\pi)^{2},
\end{equation}
For two-particle probes like optical conductivity and Raman spectroscopy one expects $p=2$. In practice the few materials where the scaling has been studied via optical conductivity have yielded $p$ values significantly different from 2 \cite{sulewski_far-infrared_1988,nagel_optical_2012,mirzaei_spectroscopic_2013,tytarenko_direct_2015,maslov_optical_2017,pustogow_rise_2021}. Deviations from the Gurzhi scaling have been attributed to energy dependent elastic scattering effects which contribute to an energy but not a temperature dependence in the quasiparticle self-energy \cite{maslov_optical_2017}. A notable exception is Sr\textsubscript{2}RuO\textsubscript{4} where Stricker et al. have reported a value very close to 2, for $\omega \leq 36$~meV and $T$ between 9 and 40~K~\cite{stricker_optical_2014}. We have attempted a similar scaling with our data by plotting $\Gamma_{\mu}$ as a function of $\omega^2+(p_{\mu}\pi k_{B} T)^2$ for $T \leq 25$~K and taking $p$ as a symmetry dependent free parameter (Fig.~\ref{fig:Gamma temperature and gurzhi}-(b) and SI~\cite{SM}). The best scalings are obtained using respectively $p \approx 1.6$ and $p \approx 0.85$ in $B_{1g}$ and $B_{2g}$, with greater uncertainties in $B_{2g}$, but indicating $p<2$ for both symmetries. Both the symmetry dependence and the discrepancy between Raman spectroscopy and optical conductivity lead us to speculate that elastic scattering effects on quasiparticle scattering rate might be orbital dependent.
\par
Overall, the symmetry dependence of the quasiparticles dynamics as probed by Raman scattering are consistent with the picture of a Hund's metal, with robust FL properties for $d_{xz/yz}$-derived quasiparticles, but more fragile for $d_{xy}$-derived ones. A remaining issue is the role of magnetic fluctuations in the strong symmetry dependence of the Raman spectra. Up to now we have essentially ignored the additional $k$ dependence embedded in the Raman vertices with the $B_{1g}$ and $B_{2g}$ vertex having line nodes at 45$\degree$ and along the principal axes respectively. In particular incommensurate antiferromagnetic fluctuations are expected to impact mostly the $d_{xz/yz}$ derived quasiparticles with hot spots located at their crossing. It is interesting to note that the $B_{1g}$ vertex precisely vanishes at these points and the $B_{2g}$ vertex has no amplitude on these orbitals. Besides, the energy dependent scattering rates suggest a gradual departure from FL behavior with a characteristic symmetry dependent crossover energy $\omega_{FL}$, without any sharp kink that could be attributed to the coupling to magnetic fluctuations (ferromagnetic and antiferromagnetic) whose characteristic energy scales lie in the energy windows studied here, below 20 meV \cite{steffens_spin_2019}. Therefore, it appears magnetic fluctuations may not play a prominent a role in the dichotomy observed in the Raman spectra.
\par 
In conclusion, we have shown that Sr\textsubscript{2}RuO\textsubscript{4} provides an attractive model system to demonstrate the ability of Raman scattering to probe orbital dependent FL properties. Our results can serve as a standard for future Raman studies aiming at probing exotic physics such as non-FL and quantum criticality. The strong dichotomy observed is in-line with the picture of Sr$_2$RuO$_4$ as a Hund's metal with coexisting robust and fragile quasiparticles. A comparison with realistic calculations of the symmetry resolved Raman spectra using for e.g. DMFT techniques should provide a more quantitative test of the Hund's metal picture. Finally, whether the observed orbital dichotomy plays a role in the superconducting pairing is an intriguing issue in light of the recent proposals of an orbital anti-symmetric chiral pairing state which was argued to reconcile several key experimental observations \cite{gingras_superconducting_2019,suh_stabilizing_2020}.

\section*{Competing interests}
The authors declare they have no competing interests

\section*{Acknowledgments}
The authors acknowledge insightful discussions with Antoine Georges, Dimitri Maslov, Indranil Paul and Yvan Sidis

\section*{Data sharing}
All relevant data are included in the main manuscript and the supplementary materials.

\section*{Author Contributions}
Y. G. designed research. J.-C. P., B. B., A. F., D. C., C. S., Y. M., M. C., A. S. and Y. G. performed research. J. -C. P. and Y. G. analyzed data. J. -C. P. and Y. G. wrote the paper.  

\section*{Funding}
J.-C. P., M. C., A. S. and Y. G. acknowledge the support from ANR Grant NEPTUN (ANR-19-CE30-0019-03)

\bibliography{apssamp}

\begin{thebibliography}{59}
\expandafter\ifx\csname natexlab\endcsname\relax\def\natexlab#1{#1}\fi
\expandafter\ifx\csname bibnamefont\endcsname\relax
  \def\bibnamefont#1{#1}\fi
\expandafter\ifx\csname bibfnamefont\endcsname\relax
  \def\bibfnamefont#1{#1}\fi
\expandafter\ifx\csname citenamefont\endcsname\relax
  \def\citenamefont#1{#1}\fi
\expandafter\ifx\csname url\endcsname\relax
  \def\url#1{\texttt{#1}}\fi
\expandafter\ifx\csname urlprefix\endcsname\relax\def\urlprefix{URL }\fi
\providecommand{\bibinfo}[2]{#2}
\providecommand{\eprint}[2][]{\url{#2}}

\bibitem[{\citenamefont{Maeno et~al.}(1994)\citenamefont{Maeno, Hashimoto,
  Yoshida, Nishizaki, Fujita, Bednorz, and
  Lichtenberg}}]{maeno_superconductivity_1994}
\bibinfo{author}{\bibfnamefont{Y.}~\bibnamefont{Maeno}},
  \bibinfo{author}{\bibfnamefont{H.}~\bibnamefont{Hashimoto}},
  \bibinfo{author}{\bibfnamefont{K.}~\bibnamefont{Yoshida}},
  \bibinfo{author}{\bibfnamefont{S.}~\bibnamefont{Nishizaki}},
  \bibinfo{author}{\bibfnamefont{T.}~\bibnamefont{Fujita}},
  \bibinfo{author}{\bibfnamefont{J.~G.} \bibnamefont{Bednorz}},
  \bibnamefont{and}
  \bibinfo{author}{\bibfnamefont{F.}~\bibnamefont{Lichtenberg}},
  \bibinfo{journal}{Nature} \textbf{\bibinfo{volume}{375}},
  \bibinfo{pages}{532} (\bibinfo{year}{1994}).

\bibitem[{\citenamefont{Mackenzie et~al.}(2017)\citenamefont{Mackenzie,
  Scaffidi, Hicks, and Maeno}}]{mackenzie_even_2017}
\bibinfo{author}{\bibfnamefont{A.~P.} \bibnamefont{Mackenzie}},
  \bibinfo{author}{\bibfnamefont{T.}~\bibnamefont{Scaffidi}},
  \bibinfo{author}{\bibfnamefont{C.~W.} \bibnamefont{Hicks}}, \bibnamefont{and}
  \bibinfo{author}{\bibfnamefont{Y.}~\bibnamefont{Maeno}},
  \bibinfo{journal}{npj Quantum Materials} \textbf{\bibinfo{volume}{2}},
  \bibinfo{pages}{40} (\bibinfo{year}{2017}), ISSN \bibinfo{issn}{2397-4648},
  \urlprefix\url{http://www.nature.com/articles/s41535-017-0045-4}.

\bibitem[{\citenamefont{Pustogow et~al.}(2019)\citenamefont{Pustogow, Luo,
  Chronister, Su, Sokolov, Jerzembeck, Mackenzie, Hicks, Kikugawa, Raghu
  et~al.}}]{pustogow_constraints_2019}
\bibinfo{author}{\bibfnamefont{A.}~\bibnamefont{Pustogow}},
  \bibinfo{author}{\bibfnamefont{Y.}~\bibnamefont{Luo}},
  \bibinfo{author}{\bibfnamefont{A.}~\bibnamefont{Chronister}},
  \bibinfo{author}{\bibfnamefont{Y.-S.} \bibnamefont{Su}},
  \bibinfo{author}{\bibfnamefont{D.~A.} \bibnamefont{Sokolov}},
  \bibinfo{author}{\bibfnamefont{F.}~\bibnamefont{Jerzembeck}},
  \bibinfo{author}{\bibfnamefont{A.~P.} \bibnamefont{Mackenzie}},
  \bibinfo{author}{\bibfnamefont{C.~W.} \bibnamefont{Hicks}},
  \bibinfo{author}{\bibfnamefont{N.}~\bibnamefont{Kikugawa}},
  \bibinfo{author}{\bibfnamefont{S.}~\bibnamefont{Raghu}},
  \bibnamefont{et~al.}, \bibinfo{journal}{Nature}
  \textbf{\bibinfo{volume}{574}}, \bibinfo{pages}{72} (\bibinfo{year}{2019}),
  ISSN \bibinfo{issn}{1476-4687},
  \urlprefix\url{https://www.nature.com/articles/s41586-019-1596-2}.

\bibitem[{\citenamefont{Bergemann et~al.}(2003)\citenamefont{Bergemann,
  Mackenzie, Julian, Forsythe, and
  Ohmichi}}]{bergemann_quasi-two-dimensional_2003}
\bibinfo{author}{\bibfnamefont{C.}~\bibnamefont{Bergemann}},
  \bibinfo{author}{\bibfnamefont{A.~P.} \bibnamefont{Mackenzie}},
  \bibinfo{author}{\bibfnamefont{S.~R.} \bibnamefont{Julian}},
  \bibinfo{author}{\bibfnamefont{D.}~\bibnamefont{Forsythe}}, \bibnamefont{and}
  \bibinfo{author}{\bibfnamefont{E.}~\bibnamefont{Ohmichi}},
  \bibinfo{journal}{Advances in Physics} \textbf{\bibinfo{volume}{52}},
  \bibinfo{pages}{639} (\bibinfo{year}{2003}), ISSN \bibinfo{issn}{0001-8732,
  1460-6976},
  \urlprefix\url{http://www.tandfonline.com/doi/abs/10.1080/00018730310001621737}.

\bibitem[{\citenamefont{Maeno et~al.}(1996)\citenamefont{Maeno, Nishizaki,
  Yoshida, Ikeda, and Fujita}}]{maeno_normal-state_1996}
\bibinfo{author}{\bibfnamefont{Y.}~\bibnamefont{Maeno}},
  \bibinfo{author}{\bibfnamefont{S.}~\bibnamefont{Nishizaki}},
  \bibinfo{author}{\bibfnamefont{K.}~\bibnamefont{Yoshida}},
  \bibinfo{author}{\bibfnamefont{S.-i.} \bibnamefont{Ikeda}}, \bibnamefont{and}
  \bibinfo{author}{\bibfnamefont{T.}~\bibnamefont{Fujita}},
  \bibinfo{journal}{Journal of Low Temperature Physics}
  \textbf{\bibinfo{volume}{105}}, \bibinfo{pages}{1577} (\bibinfo{year}{1996}).

\bibitem[{\citenamefont{Ingle et~al.}(2005)\citenamefont{Ingle, Shen,
  Baumberger, Meevasana, Lu, Shen, Damascelli, Nakatsuji, Mao, Maeno
  et~al.}}]{ingle_quantitative_2005}
\bibinfo{author}{\bibfnamefont{N.~J.~C.} \bibnamefont{Ingle}},
  \bibinfo{author}{\bibfnamefont{K.~M.} \bibnamefont{Shen}},
  \bibinfo{author}{\bibfnamefont{F.}~\bibnamefont{Baumberger}},
  \bibinfo{author}{\bibfnamefont{W.}~\bibnamefont{Meevasana}},
  \bibinfo{author}{\bibfnamefont{D.~H.} \bibnamefont{Lu}},
  \bibinfo{author}{\bibfnamefont{Z.-X.} \bibnamefont{Shen}},
  \bibinfo{author}{\bibfnamefont{A.}~\bibnamefont{Damascelli}},
  \bibinfo{author}{\bibfnamefont{S.}~\bibnamefont{Nakatsuji}},
  \bibinfo{author}{\bibfnamefont{Z.~Q.} \bibnamefont{Mao}},
  \bibinfo{author}{\bibfnamefont{Y.}~\bibnamefont{Maeno}},
  \bibnamefont{et~al.}, \bibinfo{journal}{Physical Review B}
  \textbf{\bibinfo{volume}{72}}, \bibinfo{pages}{205114}
  (\bibinfo{year}{2005}), ISSN \bibinfo{issn}{1098-0121, 1550-235X},
  \urlprefix\url{https://link.aps.org/doi/10.1103/PhysRevB.72.205114}.

\bibitem[{\citenamefont{Kidd et~al.}(2005)\citenamefont{Kidd, Valla, Fedorov,
  Johnson, Cava, and Haas}}]{kidd_orbital_2005}
\bibinfo{author}{\bibfnamefont{T.~E.} \bibnamefont{Kidd}},
  \bibinfo{author}{\bibfnamefont{T.}~\bibnamefont{Valla}},
  \bibinfo{author}{\bibfnamefont{A.~V.} \bibnamefont{Fedorov}},
  \bibinfo{author}{\bibfnamefont{P.~D.} \bibnamefont{Johnson}},
  \bibinfo{author}{\bibfnamefont{R.~J.} \bibnamefont{Cava}}, \bibnamefont{and}
  \bibinfo{author}{\bibfnamefont{M.~K.} \bibnamefont{Haas}},
  \bibinfo{journal}{Physical Review Letters} \textbf{\bibinfo{volume}{94}},
  \bibinfo{pages}{107003} (\bibinfo{year}{2005}), ISSN
  \bibinfo{issn}{0031-9007, 1079-7114},
  \urlprefix\url{https://link.aps.org/doi/10.1103/PhysRevLett.94.107003}.

\bibitem[{\citenamefont{Gurzhi}(1959)}]{gurzhi_mutual_1959}
\bibinfo{author}{\bibfnamefont{R.~N.} \bibnamefont{Gurzhi}},
  \bibinfo{journal}{Soviet Physics Journal of Experimental and Theoretical
  Physics} \textbf{\bibinfo{volume}{35 (8)}}, \bibinfo{pages}{673}
  (\bibinfo{year}{1959}).

\bibitem[{\citenamefont{Stricker et~al.}(2014)\citenamefont{Stricker, Mravlje,
  Berthod, Fittipaldi, Vecchione, Georges, and van~der
  Marel}}]{stricker_optical_2014}
\bibinfo{author}{\bibfnamefont{D.}~\bibnamefont{Stricker}},
  \bibinfo{author}{\bibfnamefont{J.}~\bibnamefont{Mravlje}},
  \bibinfo{author}{\bibfnamefont{C.}~\bibnamefont{Berthod}},
  \bibinfo{author}{\bibfnamefont{R.}~\bibnamefont{Fittipaldi}},
  \bibinfo{author}{\bibfnamefont{A.}~\bibnamefont{Vecchione}},
  \bibinfo{author}{\bibfnamefont{A.}~\bibnamefont{Georges}}, \bibnamefont{and}
  \bibinfo{author}{\bibfnamefont{D.}~\bibnamefont{van~der Marel}},
  \bibinfo{journal}{Physical Review Letters} \textbf{\bibinfo{volume}{113}},
  \bibinfo{pages}{087404} (\bibinfo{year}{2014}).

\bibitem[{\citenamefont{Maslov and Chubukov}(2017)}]{maslov_optical_2017}
\bibinfo{author}{\bibfnamefont{D.~L.} \bibnamefont{Maslov}} \bibnamefont{and}
  \bibinfo{author}{\bibfnamefont{A.~V.} \bibnamefont{Chubukov}},
  \bibinfo{journal}{Reports on Progress in Physics}
  \textbf{\bibinfo{volume}{80}}, \bibinfo{pages}{026503}
  (\bibinfo{year}{2017}), ISSN \bibinfo{issn}{0034-4885, 1361-6633},
  \urlprefix\url{https://iopscience.iop.org/article/10.1088/1361-6633/80/2/026503}.

\bibitem[{\citenamefont{Maeno et~al.}(1997)\citenamefont{Maeno, Yoshida,
  Hashimoto, Nishizaki, Ikeda, Nohara, Fujita, Mackenzie, Hussey, Bednorz
  et~al.}}]{maeno_two-dimensional_1997}
\bibinfo{author}{\bibfnamefont{Y.}~\bibnamefont{Maeno}},
  \bibinfo{author}{\bibfnamefont{K.}~\bibnamefont{Yoshida}},
  \bibinfo{author}{\bibfnamefont{H.}~\bibnamefont{Hashimoto}},
  \bibinfo{author}{\bibfnamefont{S.}~\bibnamefont{Nishizaki}},
  \bibinfo{author}{\bibfnamefont{S.-i.} \bibnamefont{Ikeda}},
  \bibinfo{author}{\bibfnamefont{M.}~\bibnamefont{Nohara}},
  \bibinfo{author}{\bibfnamefont{T.}~\bibnamefont{Fujita}},
  \bibinfo{author}{\bibfnamefont{A.~P.} \bibnamefont{Mackenzie}},
  \bibinfo{author}{\bibfnamefont{N.~E.} \bibnamefont{Hussey}},
  \bibinfo{author}{\bibfnamefont{J.~G.} \bibnamefont{Bednorz}},
  \bibnamefont{et~al.}, \bibinfo{journal}{Journal of the Physical Society of
  Japan} \textbf{\bibinfo{volume}{66}}, \bibinfo{pages}{1405}
  (\bibinfo{year}{1997}).

\bibitem[{\citenamefont{Mackenzie et~al.}(1998)\citenamefont{Mackenzie, Ikeda,
  Maeno, Fujita, Julian, and Lonzarich}}]{mackenzie_fermi_1998}
\bibinfo{author}{\bibfnamefont{A.}~\bibnamefont{Mackenzie}},
  \bibinfo{author}{\bibfnamefont{S.-i.} \bibnamefont{Ikeda}},
  \bibinfo{author}{\bibfnamefont{Y.}~\bibnamefont{Maeno}},
  \bibinfo{author}{\bibfnamefont{T.}~\bibnamefont{Fujita}},
  \bibinfo{author}{\bibfnamefont{S.}~\bibnamefont{Julian}}, \bibnamefont{and}
  \bibinfo{author}{\bibfnamefont{G.}~\bibnamefont{Lonzarich}},
  \bibinfo{journal}{Journal of the Physical Society of Japan}
  \textbf{\bibinfo{volume}{67}}, \bibinfo{pages}{385} (\bibinfo{year}{1998}),
  ISSN \bibinfo{issn}{0031-9015, 1347-4073},
  \urlprefix\url{http://journals.jps.jp/doi/10.1143/JPSJ.67.385}.

\bibitem[{\citenamefont{Tamai et~al.}(2019)\citenamefont{Tamai, Zingl,
  Rozbicki, Cappelli, Ricco, de~la Torre, McKeown~Walker, Bruno, King,
  Meevasana et~al.}}]{tamai_high-resolution_2019}
\bibinfo{author}{\bibfnamefont{A.}~\bibnamefont{Tamai}},
  \bibinfo{author}{\bibfnamefont{M.}~\bibnamefont{Zingl}},
  \bibinfo{author}{\bibfnamefont{E.}~\bibnamefont{Rozbicki}},
  \bibinfo{author}{\bibfnamefont{E.}~\bibnamefont{Cappelli}},
  \bibinfo{author}{\bibfnamefont{S.}~\bibnamefont{Ricco}},
  \bibinfo{author}{\bibfnamefont{A.}~\bibnamefont{de~la Torre}},
  \bibinfo{author}{\bibfnamefont{S.}~\bibnamefont{McKeown~Walker}},
  \bibinfo{author}{\bibfnamefont{F.~Y.} \bibnamefont{Bruno}},
  \bibinfo{author}{\bibfnamefont{P.~D.~C.} \bibnamefont{King}},
  \bibinfo{author}{\bibfnamefont{W.}~\bibnamefont{Meevasana}},
  \bibnamefont{et~al.}, \bibinfo{journal}{Physical Review X}
  \textbf{\bibinfo{volume}{9}}, \bibinfo{pages}{021048} (\bibinfo{year}{2019}).

\bibitem[{\citenamefont{Katsufuji et~al.}(1996)\citenamefont{Katsufuji, Kasai,
  and Tokura}}]{katsufuji_in-plane_1996}
\bibinfo{author}{\bibfnamefont{T.}~\bibnamefont{Katsufuji}},
  \bibinfo{author}{\bibfnamefont{M.}~\bibnamefont{Kasai}}, \bibnamefont{and}
  \bibinfo{author}{\bibfnamefont{Y.}~\bibnamefont{Tokura}},
  \bibinfo{journal}{Physical Review Letters} \textbf{\bibinfo{volume}{76}},
  \bibinfo{pages}{126} (\bibinfo{year}{1996}).

\bibitem[{\citenamefont{Lee et~al.}(2006)\citenamefont{Lee, Moon, Noh,
  Nakatsuji, and Maeno}}]{lee_orbital-selective_2006}
\bibinfo{author}{\bibfnamefont{J.~S.} \bibnamefont{Lee}},
  \bibinfo{author}{\bibfnamefont{S.~J.} \bibnamefont{Moon}},
  \bibinfo{author}{\bibfnamefont{T.~W.} \bibnamefont{Noh}},
  \bibinfo{author}{\bibfnamefont{S.}~\bibnamefont{Nakatsuji}},
  \bibnamefont{and} \bibinfo{author}{\bibfnamefont{Y.}~\bibnamefont{Maeno}},
  \bibinfo{journal}{Physical Review Letters} \textbf{\bibinfo{volume}{96}},
  \bibinfo{pages}{057401} (\bibinfo{year}{2006}), ISSN
  \bibinfo{issn}{0031-9007, 1079-7114},
  \urlprefix\url{https://link.aps.org/doi/10.1103/PhysRevLett.96.057401}.

\bibitem[{\citenamefont{Mravlje et~al.}(2011)\citenamefont{Mravlje, Aichhorn,
  Miyake, Haule, Kotliar, and Georges}}]{mravlje_coherence-incoherence_2011}
\bibinfo{author}{\bibfnamefont{J.}~\bibnamefont{Mravlje}},
  \bibinfo{author}{\bibfnamefont{M.}~\bibnamefont{Aichhorn}},
  \bibinfo{author}{\bibfnamefont{T.}~\bibnamefont{Miyake}},
  \bibinfo{author}{\bibfnamefont{K.}~\bibnamefont{Haule}},
  \bibinfo{author}{\bibfnamefont{G.}~\bibnamefont{Kotliar}}, \bibnamefont{and}
  \bibinfo{author}{\bibfnamefont{A.}~\bibnamefont{Georges}},
  \bibinfo{journal}{Physical Review Letters} \textbf{\bibinfo{volume}{106}},
  \bibinfo{pages}{096401} (\bibinfo{year}{2011}).

\bibitem[{\citenamefont{Georges et~al.}(2013)\citenamefont{Georges, Medici, and
  Mravlje}}]{georges_strong_2013}
\bibinfo{author}{\bibfnamefont{A.}~\bibnamefont{Georges}},
  \bibinfo{author}{\bibfnamefont{L.~d.} \bibnamefont{Medici}},
  \bibnamefont{and} \bibinfo{author}{\bibfnamefont{J.}~\bibnamefont{Mravlje}},
  \bibinfo{journal}{Annual Review of Condensed Matter Physics}
  \textbf{\bibinfo{volume}{4}}, \bibinfo{pages}{137} (\bibinfo{year}{2013}),
  ISSN \bibinfo{issn}{1947-5454, 1947-5462},
  \urlprefix\url{http://www.annualreviews.org/doi/10.1146/annurev-conmatphys-020911-125045}.

\bibitem[{\citenamefont{Mackenzie et~al.}(1996)\citenamefont{Mackenzie, Julian,
  Diver, McMullan, Ray, Lonzarich, Maeno, Nishizaki, and
  Fujita}}]{mackenzie_quantum_1996}
\bibinfo{author}{\bibfnamefont{A.~P.} \bibnamefont{Mackenzie}},
  \bibinfo{author}{\bibfnamefont{S.~R.} \bibnamefont{Julian}},
  \bibinfo{author}{\bibfnamefont{A.~J.} \bibnamefont{Diver}},
  \bibinfo{author}{\bibfnamefont{G.~J.} \bibnamefont{McMullan}},
  \bibinfo{author}{\bibfnamefont{M.~P.} \bibnamefont{Ray}},
  \bibinfo{author}{\bibfnamefont{G.~G.} \bibnamefont{Lonzarich}},
  \bibinfo{author}{\bibfnamefont{Y.}~\bibnamefont{Maeno}},
  \bibinfo{author}{\bibfnamefont{S.}~\bibnamefont{Nishizaki}},
  \bibnamefont{and} \bibinfo{author}{\bibfnamefont{T.}~\bibnamefont{Fujita}},
  \bibinfo{journal}{Physical Review Letters} \textbf{\bibinfo{volume}{76}},
  \bibinfo{pages}{3786} (\bibinfo{year}{1996}).

\bibitem[{\citenamefont{Bergemann et~al.}(2000)\citenamefont{Bergemann, Julian,
  Mackenzie, Nishizaki, and Maeno}}]{bergemann_detailed_2000}
\bibinfo{author}{\bibfnamefont{C.}~\bibnamefont{Bergemann}},
  \bibinfo{author}{\bibfnamefont{S.~R.} \bibnamefont{Julian}},
  \bibinfo{author}{\bibfnamefont{A.~P.} \bibnamefont{Mackenzie}},
  \bibinfo{author}{\bibfnamefont{S.}~\bibnamefont{Nishizaki}},
  \bibnamefont{and} \bibinfo{author}{\bibfnamefont{Y.}~\bibnamefont{Maeno}},
  \bibinfo{journal}{Physical Review Letters} \textbf{\bibinfo{volume}{84}},
  \bibinfo{pages}{2662} (\bibinfo{year}{2000}).

\bibitem[{\citenamefont{Damascelli et~al.}(2000)\citenamefont{Damascelli, Lu,
  Shen, Armitage, Ronning, Feng, Kim, Shen, Kimura, Tokura
  et~al.}}]{damascelli_fermi_2000}
\bibinfo{author}{\bibfnamefont{A.}~\bibnamefont{Damascelli}},
  \bibinfo{author}{\bibfnamefont{D.~H.} \bibnamefont{Lu}},
  \bibinfo{author}{\bibfnamefont{K.~M.} \bibnamefont{Shen}},
  \bibinfo{author}{\bibfnamefont{N.~P.} \bibnamefont{Armitage}},
  \bibinfo{author}{\bibfnamefont{F.}~\bibnamefont{Ronning}},
  \bibinfo{author}{\bibfnamefont{D.~L.} \bibnamefont{Feng}},
  \bibinfo{author}{\bibfnamefont{C.}~\bibnamefont{Kim}},
  \bibinfo{author}{\bibfnamefont{Z.-X.} \bibnamefont{Shen}},
  \bibinfo{author}{\bibfnamefont{T.}~\bibnamefont{Kimura}},
  \bibinfo{author}{\bibfnamefont{Y.}~\bibnamefont{Tokura}},
  \bibnamefont{et~al.}, \bibinfo{journal}{Physical Review Letters}
  \textbf{\bibinfo{volume}{85}}, \bibinfo{pages}{5194} (\bibinfo{year}{2000}).

\bibitem[{\citenamefont{Kugler et~al.}(2020)\citenamefont{Kugler, Zingl,
  Strand, Lee, von Delft, and Georges}}]{kugler_strongly_2020}
\bibinfo{author}{\bibfnamefont{F.~B.} \bibnamefont{Kugler}},
  \bibinfo{author}{\bibfnamefont{M.}~\bibnamefont{Zingl}},
  \bibinfo{author}{\bibfnamefont{H.~U.} \bibnamefont{Strand}},
  \bibinfo{author}{\bibfnamefont{S.-S.~B.} \bibnamefont{Lee}},
  \bibinfo{author}{\bibfnamefont{J.}~\bibnamefont{von Delft}},
  \bibnamefont{and} \bibinfo{author}{\bibfnamefont{A.}~\bibnamefont{Georges}},
  \bibinfo{journal}{Physical Review Letters} \textbf{\bibinfo{volume}{124}},
  \bibinfo{pages}{016401} (\bibinfo{year}{2020}), ISSN
  \bibinfo{issn}{0031-9007, 1079-7114},
  \urlprefix\url{https://link.aps.org/doi/10.1103/PhysRevLett.124.016401}.

\bibitem[{\citenamefont{Mazin and Singh}(1999)}]{mazin_competitions_1999}
\bibinfo{author}{\bibfnamefont{I.~I.} \bibnamefont{Mazin}} \bibnamefont{and}
  \bibinfo{author}{\bibfnamefont{D.~J.} \bibnamefont{Singh}},
  \bibinfo{journal}{Physical Review Letters} \textbf{\bibinfo{volume}{82}},
  \bibinfo{pages}{4324} (\bibinfo{year}{1999}),
  \urlprefix\url{https://link.aps.org/doi/10.1103/PhysRevLett.82.4324}.

\bibitem[{\citenamefont{Rømer et~al.}(2019)\citenamefont{Rømer, Scherer,
  Eremin, Hirschfeld, and Andersen}}]{romer_knight_2019}
\bibinfo{author}{\bibfnamefont{A.}~\bibnamefont{Rømer}},
  \bibinfo{author}{\bibfnamefont{D.}~\bibnamefont{Scherer}},
  \bibinfo{author}{\bibfnamefont{I.}~\bibnamefont{Eremin}},
  \bibinfo{author}{\bibfnamefont{P.}~\bibnamefont{Hirschfeld}},
  \bibnamefont{and} \bibinfo{author}{\bibfnamefont{B.}~\bibnamefont{Andersen}},
  \bibinfo{journal}{Physical Review Letters} \textbf{\bibinfo{volume}{123}},
  \bibinfo{pages}{247001} (\bibinfo{year}{2019}), ISSN
  \bibinfo{issn}{0031-9007, 1079-7114},
  \urlprefix\url{https://link.aps.org/doi/10.1103/PhysRevLett.123.247001}.

\bibitem[{\citenamefont{Sidis et~al.}(1999)\citenamefont{Sidis, Braden,
  Bourges, Hennion, NishiZaki, Maeno, and Mori}}]{sidis_evidence_1999}
\bibinfo{author}{\bibfnamefont{Y.}~\bibnamefont{Sidis}},
  \bibinfo{author}{\bibfnamefont{M.}~\bibnamefont{Braden}},
  \bibinfo{author}{\bibfnamefont{P.}~\bibnamefont{Bourges}},
  \bibinfo{author}{\bibfnamefont{B.}~\bibnamefont{Hennion}},
  \bibinfo{author}{\bibfnamefont{S.}~\bibnamefont{NishiZaki}},
  \bibinfo{author}{\bibfnamefont{Y.}~\bibnamefont{Maeno}}, \bibnamefont{and}
  \bibinfo{author}{\bibfnamefont{Y.}~\bibnamefont{Mori}},
  \bibinfo{journal}{Physical Review Letters} \textbf{\bibinfo{volume}{83}},
  \bibinfo{pages}{3320} (\bibinfo{year}{1999}), ISSN \bibinfo{issn}{0031-9007,
  1079-7114},
  \urlprefix\url{https://link.aps.org/doi/10.1103/PhysRevLett.83.3320}.

\bibitem[{\citenamefont{Steffens et~al.}(2019)\citenamefont{Steffens, Sidis,
  Kulda, Mao, Maeno, Mazin, and Braden}}]{steffens_spin_2019}
\bibinfo{author}{\bibfnamefont{P.}~\bibnamefont{Steffens}},
  \bibinfo{author}{\bibfnamefont{Y.}~\bibnamefont{Sidis}},
  \bibinfo{author}{\bibfnamefont{J.}~\bibnamefont{Kulda}},
  \bibinfo{author}{\bibfnamefont{Z.}~\bibnamefont{Mao}},
  \bibinfo{author}{\bibfnamefont{Y.}~\bibnamefont{Maeno}},
  \bibinfo{author}{\bibfnamefont{I.}~\bibnamefont{Mazin}}, \bibnamefont{and}
  \bibinfo{author}{\bibfnamefont{M.}~\bibnamefont{Braden}},
  \bibinfo{journal}{Physical Review Letters} \textbf{\bibinfo{volume}{122}},
  \bibinfo{pages}{047004} (\bibinfo{year}{2019}), ISSN
  \bibinfo{issn}{0031-9007, 1079-7114},
  \urlprefix\url{https://link.aps.org/doi/10.1103/PhysRevLett.122.047004}.

\bibitem[{SM()}]{SM}
 (????), \bibinfo{note}{supplementary Material to this paper includes an
  estimation of laser heating, additional spectra at higher temperatures, in
  the ${E}_{g}$ symmetry channel and at different laser wavelengths, a
  description of the inversion procedure of the extended {D}rude model, details
  on the {R}aman vertex calculations, additional study on the quadratic
  behavior of the scattering rate as a function of frequency and temperature,
  additional details on the determination of $p_{\mu}$}.

\bibitem[{\citenamefont{Bobowski et~al.}(2019)\citenamefont{Bobowski, Kikugawa,
  Miyoshi, Suwa, Xu, Yonezawa, Sokolov, Mackenzie, and
  Maeno}}]{bobowski_improved_2019}
\bibinfo{author}{\bibfnamefont{J.~S.} \bibnamefont{Bobowski}},
  \bibinfo{author}{\bibfnamefont{N.}~\bibnamefont{Kikugawa}},
  \bibinfo{author}{\bibfnamefont{T.}~\bibnamefont{Miyoshi}},
  \bibinfo{author}{\bibfnamefont{H.}~\bibnamefont{Suwa}},
  \bibinfo{author}{\bibfnamefont{H.-s.} \bibnamefont{Xu}},
  \bibinfo{author}{\bibfnamefont{S.}~\bibnamefont{Yonezawa}},
  \bibinfo{author}{\bibfnamefont{D.~A.} \bibnamefont{Sokolov}},
  \bibinfo{author}{\bibfnamefont{A.~P.} \bibnamefont{Mackenzie}},
  \bibnamefont{and} \bibinfo{author}{\bibfnamefont{Y.}~\bibnamefont{Maeno}},
  \bibinfo{journal}{Condensed Matter} \textbf{\bibinfo{volume}{4}},
  \bibinfo{pages}{6} (\bibinfo{year}{2019}), ISSN \bibinfo{issn}{2410-3896},
  \urlprefix\url{https://www.mdpi.com/2410-3896/4/1/6}.

\bibitem[{\citenamefont{Devereaux and Hackl}(2007)}]{devereaux_inelastic_2007}
\bibinfo{author}{\bibfnamefont{T.~P.} \bibnamefont{Devereaux}}
  \bibnamefont{and} \bibinfo{author}{\bibfnamefont{R.}~\bibnamefont{Hackl}},
  \bibinfo{journal}{Reviews of Modern Physics} \textbf{\bibinfo{volume}{79}},
  \bibinfo{pages}{175} (\bibinfo{year}{2007}),
  \urlprefix\url{https://link.aps.org/doi/10.1103/RevModPhys.79.175}.

\bibitem[{\citenamefont{Cobo et~al.}(2016)\citenamefont{Cobo, Ahn, Eremin, and
  Akbari}}]{cobo_anisotropic_2016}
\bibinfo{author}{\bibfnamefont{S.}~\bibnamefont{Cobo}},
  \bibinfo{author}{\bibfnamefont{F.}~\bibnamefont{Ahn}},
  \bibinfo{author}{\bibfnamefont{I.}~\bibnamefont{Eremin}}, \bibnamefont{and}
  \bibinfo{author}{\bibfnamefont{A.}~\bibnamefont{Akbari}},
  \bibinfo{journal}{Physical Review B} \textbf{\bibinfo{volume}{94}},
  \bibinfo{pages}{224507} (\bibinfo{year}{2016}), ISSN
  \bibinfo{issn}{2469-9950, 2469-9969},
  \urlprefix\url{https://link.aps.org/doi/10.1103/PhysRevB.94.224507}.

\bibitem[{\citenamefont{Ipatova et~al.}(1981)\citenamefont{Ipatova, Subashiev,
  and Voitenko}}]{ipatova_electron_nodate}
\bibinfo{author}{\bibfnamefont{I.~P.} \bibnamefont{Ipatova}},
  \bibinfo{author}{\bibfnamefont{A.~V.} \bibnamefont{Subashiev}},
  \bibnamefont{and} \bibinfo{author}{\bibfnamefont{V.~A.}
  \bibnamefont{Voitenko}}, \bibinfo{journal}{Solid State Communications}
  \textbf{\bibinfo{volume}{37}}, \bibinfo{pages}{893} (\bibinfo{year}{1981}).

\bibitem[{\citenamefont{Zawadowski and Cardona}(1990)}]{zawadowski_theory_1990}
\bibinfo{author}{\bibfnamefont{A.}~\bibnamefont{Zawadowski}} \bibnamefont{and}
  \bibinfo{author}{\bibfnamefont{M.}~\bibnamefont{Cardona}},
  \bibinfo{journal}{Physical Review B} \textbf{\bibinfo{volume}{42}},
  \bibinfo{pages}{10732} (\bibinfo{year}{1990}), ISSN \bibinfo{issn}{0163-1829,
  1095-3795},
  \urlprefix\url{https://link.aps.org/doi/10.1103/PhysRevB.42.10732}.

\bibitem[{\citenamefont{Berthod et~al.}(2013)\citenamefont{Berthod, Mravlje,
  Deng, Žitko, van~der Marel, and Georges}}]{berthod_non-drude_2013}
\bibinfo{author}{\bibfnamefont{C.}~\bibnamefont{Berthod}},
  \bibinfo{author}{\bibfnamefont{J.}~\bibnamefont{Mravlje}},
  \bibinfo{author}{\bibfnamefont{X.}~\bibnamefont{Deng}},
  \bibinfo{author}{\bibfnamefont{R.}~\bibnamefont{Žitko}},
  \bibinfo{author}{\bibfnamefont{D.}~\bibnamefont{van~der Marel}},
  \bibnamefont{and} \bibinfo{author}{\bibfnamefont{A.}~\bibnamefont{Georges}},
  \bibinfo{journal}{Physical Review B} \textbf{\bibinfo{volume}{87}},
  \bibinfo{pages}{115109} (\bibinfo{year}{2013}),
  \urlprefix\url{https://link.aps.org/doi/10.1103/PhysRevB.87.115109}.

\bibitem[{\citenamefont{Opel et~al.}(2000)\citenamefont{Opel, Nemetschek,
  Hoffmann, Philipp, Müller, Hackl, Tüttő, Erb, Revaz, Walker
  et~al.}}]{opel_carrier_2000}
\bibinfo{author}{\bibfnamefont{M.}~\bibnamefont{Opel}},
  \bibinfo{author}{\bibfnamefont{R.}~\bibnamefont{Nemetschek}},
  \bibinfo{author}{\bibfnamefont{C.}~\bibnamefont{Hoffmann}},
  \bibinfo{author}{\bibfnamefont{R.}~\bibnamefont{Philipp}},
  \bibinfo{author}{\bibfnamefont{P.~F.} \bibnamefont{Müller}},
  \bibinfo{author}{\bibfnamefont{R.}~\bibnamefont{Hackl}},
  \bibinfo{author}{\bibfnamefont{I.}~\bibnamefont{Tüttő}},
  \bibinfo{author}{\bibfnamefont{A.}~\bibnamefont{Erb}},
  \bibinfo{author}{\bibfnamefont{B.}~\bibnamefont{Revaz}},
  \bibinfo{author}{\bibfnamefont{E.}~\bibnamefont{Walker}},
  \bibnamefont{et~al.}, \bibinfo{journal}{Physical Review B}
  \textbf{\bibinfo{volume}{61}}, \bibinfo{pages}{9752} (\bibinfo{year}{2000}),
  ISSN \bibinfo{issn}{0163-1829, 1095-3795},
  \urlprefix\url{https://link.aps.org/doi/10.1103/PhysRevB.61.9752}.

\bibitem[{\citenamefont{Sen et~al.}(2020)\citenamefont{Sen, Fuchs, Heid,
  Kleindienst, Wolff, Schmalian, and Le~Tacon}}]{sen_strange_2020}
\bibinfo{author}{\bibfnamefont{K.}~\bibnamefont{Sen}},
  \bibinfo{author}{\bibfnamefont{D.}~\bibnamefont{Fuchs}},
  \bibinfo{author}{\bibfnamefont{R.}~\bibnamefont{Heid}},
  \bibinfo{author}{\bibfnamefont{K.}~\bibnamefont{Kleindienst}},
  \bibinfo{author}{\bibfnamefont{K.}~\bibnamefont{Wolff}},
  \bibinfo{author}{\bibfnamefont{J.}~\bibnamefont{Schmalian}},
  \bibnamefont{and} \bibinfo{author}{\bibfnamefont{M.}~\bibnamefont{Le~Tacon}},
  \bibinfo{journal}{Nature Communications} \textbf{\bibinfo{volume}{11}},
  \bibinfo{pages}{4270} (\bibinfo{year}{2020}), ISSN \bibinfo{issn}{2041-1723},
  \urlprefix\url{https://www.nature.com/articles/s41467-020-18092-6}.

\bibitem[{\citenamefont{Udagawa et~al.}(1996)\citenamefont{Udagawa, Minami,
  Ogita, Maeno, Nakamura, Fujita, Bednorz, and
  Lichtenberg}}]{udagawa_phonon_1996}
\bibinfo{author}{\bibfnamefont{M.}~\bibnamefont{Udagawa}},
  \bibinfo{author}{\bibfnamefont{T.}~\bibnamefont{Minami}},
  \bibinfo{author}{\bibfnamefont{N.}~\bibnamefont{Ogita}},
  \bibinfo{author}{\bibfnamefont{Y.}~\bibnamefont{Maeno}},
  \bibinfo{author}{\bibfnamefont{F.}~\bibnamefont{Nakamura}},
  \bibinfo{author}{\bibfnamefont{T.}~\bibnamefont{Fujita}},
  \bibinfo{author}{\bibfnamefont{J.~G.} \bibnamefont{Bednorz}},
  \bibnamefont{and}
  \bibinfo{author}{\bibfnamefont{F.}~\bibnamefont{Lichtenberg}},
  \bibinfo{journal}{Physica B: Condensed Matter}
  \textbf{\bibinfo{volume}{219-220}}, \bibinfo{pages}{222}
  (\bibinfo{year}{1996}), ISSN \bibinfo{issn}{09214526},
  \urlprefix\url{https://linkinghub.elsevier.com/retrieve/pii/0921452695007024}.

\bibitem[{\citenamefont{Sakita et~al.}(2001)\citenamefont{Sakita, Nimori, Mao,
  Maeno, Ogita, and Udagawa}}]{sakita_anisotropic_2001}
\bibinfo{author}{\bibfnamefont{S.}~\bibnamefont{Sakita}},
  \bibinfo{author}{\bibfnamefont{S.}~\bibnamefont{Nimori}},
  \bibinfo{author}{\bibfnamefont{Z.~Q.} \bibnamefont{Mao}},
  \bibinfo{author}{\bibfnamefont{Y.}~\bibnamefont{Maeno}},
  \bibinfo{author}{\bibfnamefont{N.}~\bibnamefont{Ogita}}, \bibnamefont{and}
  \bibinfo{author}{\bibfnamefont{M.}~\bibnamefont{Udagawa}},
  \bibinfo{journal}{Physical Review B} \textbf{\bibinfo{volume}{63}},
  \bibinfo{pages}{134520} (\bibinfo{year}{2001}).

\bibitem[{\citenamefont{Iliev et~al.}(2005)\citenamefont{Iliev, Popov,
  Litvinchuk, Abrashev, Bäckström, Sun, Meng, and
  Chu}}]{iliev_comparative_2005}
\bibinfo{author}{\bibfnamefont{M.}~\bibnamefont{Iliev}},
  \bibinfo{author}{\bibfnamefont{V.}~\bibnamefont{Popov}},
  \bibinfo{author}{\bibfnamefont{A.}~\bibnamefont{Litvinchuk}},
  \bibinfo{author}{\bibfnamefont{M.}~\bibnamefont{Abrashev}},
  \bibinfo{author}{\bibfnamefont{J.}~\bibnamefont{Bäckström}},
  \bibinfo{author}{\bibfnamefont{Y.}~\bibnamefont{Sun}},
  \bibinfo{author}{\bibfnamefont{R.}~\bibnamefont{Meng}}, \bibnamefont{and}
  \bibinfo{author}{\bibfnamefont{C.}~\bibnamefont{Chu}},
  \bibinfo{journal}{Physica B: Condensed Matter}
  \textbf{\bibinfo{volume}{358}}, \bibinfo{pages}{138} (\bibinfo{year}{2005}),
  ISSN \bibinfo{issn}{09214526},
  \urlprefix\url{https://linkinghub.elsevier.com/retrieve/pii/S0921452605000037}.

\bibitem[{\citenamefont{Hussey et~al.}(1998)\citenamefont{Hussey, Mackenzie,
  Cooper, Maeno, Nishizaki, and Fujita}}]{hussey_normal-state_1998}
\bibinfo{author}{\bibfnamefont{N.~E.} \bibnamefont{Hussey}},
  \bibinfo{author}{\bibfnamefont{A.~P.} \bibnamefont{Mackenzie}},
  \bibinfo{author}{\bibfnamefont{J.~R.} \bibnamefont{Cooper}},
  \bibinfo{author}{\bibfnamefont{Y.}~\bibnamefont{Maeno}},
  \bibinfo{author}{\bibfnamefont{S.}~\bibnamefont{Nishizaki}},
  \bibnamefont{and} \bibinfo{author}{\bibfnamefont{T.}~\bibnamefont{Fujita}},
  \bibinfo{journal}{Physical Review B} \textbf{\bibinfo{volume}{57}},
  \bibinfo{pages}{5505} (\bibinfo{year}{1998}),
  \urlprefix\url{https://link.aps.org/doi/10.1103/PhysRevB.57.5505}.

\bibitem[{\citenamefont{Husain et~al.}(2020)\citenamefont{Husain, Mitrano, Rak,
  Rubeck, Yang, Sow, Maeno, Batson, and Abbamonte}}]{husain_coexisting_2020}
\bibinfo{author}{\bibfnamefont{A.~A.} \bibnamefont{Husain}},
  \bibinfo{author}{\bibfnamefont{M.}~\bibnamefont{Mitrano}},
  \bibinfo{author}{\bibfnamefont{M.~S.} \bibnamefont{Rak}},
  \bibinfo{author}{\bibfnamefont{S.~I.} \bibnamefont{Rubeck}},
  \bibinfo{author}{\bibfnamefont{H.}~\bibnamefont{Yang}},
  \bibinfo{author}{\bibfnamefont{C.}~\bibnamefont{Sow}},
  \bibinfo{author}{\bibfnamefont{Y.}~\bibnamefont{Maeno}},
  \bibinfo{author}{\bibfnamefont{P.~E.} \bibnamefont{Batson}},
  \bibnamefont{and}
  \bibinfo{author}{\bibfnamefont{P.}~\bibnamefont{Abbamonte}},
  \bibinfo{journal}{arXiv:2007.06670 [cond-mat]}  (\bibinfo{year}{2020}),
  \urlprefix\url{http://arxiv.org/abs/2007.06670}.

\bibitem[{\citenamefont{Götze and Wölfle}(1972)}]{gotze_homogeneous_1972}
\bibinfo{author}{\bibfnamefont{W.}~\bibnamefont{Götze}} \bibnamefont{and}
  \bibinfo{author}{\bibfnamefont{P.}~\bibnamefont{Wölfle}},
  \bibinfo{journal}{Physical Review B} \textbf{\bibinfo{volume}{6}},
  \bibinfo{pages}{1226} (\bibinfo{year}{1972}), ISSN \bibinfo{issn}{0556-2805},
  \urlprefix\url{https://link.aps.org/doi/10.1103/PhysRevB.6.1226}.

\bibitem[{\citenamefont{McMullan et~al.}(1996)\citenamefont{McMullan, Ray, and
  Needs}}]{mcmullan_comparison_1996}
\bibinfo{author}{\bibfnamefont{G.~J.} \bibnamefont{McMullan}},
  \bibinfo{author}{\bibfnamefont{M.~P.} \bibnamefont{Ray}}, \bibnamefont{and}
  \bibinfo{author}{\bibfnamefont{R.~J.} \bibnamefont{Needs}},
  \bibinfo{journal}{Physica B} \textbf{\bibinfo{volume}{223-224}},
  \bibinfo{pages}{529} (\bibinfo{year}{1996}).

\bibitem[{\citenamefont{Veenstra et~al.}(2014)\citenamefont{Veenstra, Zhu,
  Raichle, Ludbrook, Nicolaou, Slomski, Landolt, Kittaka, Maeno, Dil
  et~al.}}]{veenstra_spin-orbital_2014}
\bibinfo{author}{\bibfnamefont{C.}~\bibnamefont{Veenstra}},
  \bibinfo{author}{\bibfnamefont{Z.-H.} \bibnamefont{Zhu}},
  \bibinfo{author}{\bibfnamefont{M.}~\bibnamefont{Raichle}},
  \bibinfo{author}{\bibfnamefont{B.}~\bibnamefont{Ludbrook}},
  \bibinfo{author}{\bibfnamefont{A.}~\bibnamefont{Nicolaou}},
  \bibinfo{author}{\bibfnamefont{B.}~\bibnamefont{Slomski}},
  \bibinfo{author}{\bibfnamefont{G.}~\bibnamefont{Landolt}},
  \bibinfo{author}{\bibfnamefont{S.}~\bibnamefont{Kittaka}},
  \bibinfo{author}{\bibfnamefont{Y.}~\bibnamefont{Maeno}},
  \bibinfo{author}{\bibfnamefont{J.}~\bibnamefont{Dil}}, \bibnamefont{et~al.},
  \bibinfo{journal}{Physical Review Letters} \textbf{\bibinfo{volume}{112}},
  \bibinfo{pages}{127002} (\bibinfo{year}{2014}), ISSN
  \bibinfo{issn}{0031-9007, 1079-7114},
  \urlprefix\url{https://link.aps.org/doi/10.1103/PhysRevLett.112.127002}.

\bibitem[{\citenamefont{Rice}(1968)}]{rice_electron-electron_1968}
\bibinfo{author}{\bibfnamefont{M.~J.} \bibnamefont{Rice}},
  \bibinfo{journal}{Physical Review Letters} \textbf{\bibinfo{volume}{20}},
  \bibinfo{pages}{1439} (\bibinfo{year}{1968}), ISSN \bibinfo{issn}{0031-9007},
  \urlprefix\url{https://link.aps.org/doi/10.1103/PhysRevLett.20.1439}.

\bibitem[{\citenamefont{Kadowaki and Woods}(1986)}]{noauthor_universal_1986}
\bibinfo{author}{\bibfnamefont{K.}~\bibnamefont{Kadowaki}} \bibnamefont{and}
  \bibinfo{author}{\bibfnamefont{S.~B.} \bibnamefont{Woods}},
  \bibinfo{journal}{Solid State Communications} \textbf{\bibinfo{volume}{58}},
  \bibinfo{pages}{507} (\bibinfo{year}{1986}), ISSN \bibinfo{issn}{0038-1098},
  \urlprefix\url{http://www.sciencedirect.com/science/article/pii/0038109886907854}.

\bibitem[{\citenamefont{Hussey}(2005)}]{hussey_non-generality_2005}
\bibinfo{author}{\bibfnamefont{N.~E.} \bibnamefont{Hussey}},
  \bibinfo{journal}{Journal of the Physical Society of Japan}
  \textbf{\bibinfo{volume}{74}}, \bibinfo{pages}{1107} (\bibinfo{year}{2005}),
  ISSN \bibinfo{issn}{0031-9015, 1347-4073},
  \urlprefix\url{http://arxiv.org/abs/cond-mat/0409252}.

\bibitem[{\citenamefont{Jacko et~al.}(2009)\citenamefont{Jacko, Fjærestad, and
  Powell}}]{jacko_unified_2009}
\bibinfo{author}{\bibfnamefont{A.~C.} \bibnamefont{Jacko}},
  \bibinfo{author}{\bibfnamefont{J.~O.} \bibnamefont{Fjærestad}},
  \bibnamefont{and} \bibinfo{author}{\bibfnamefont{B.~J.}
  \bibnamefont{Powell}}, \bibinfo{journal}{Nature Physics}
  \textbf{\bibinfo{volume}{5}}, \bibinfo{pages}{422} (\bibinfo{year}{2009}),
  ISSN \bibinfo{issn}{1745-2481},
  \urlprefix\url{https://www.nature.com/articles/nphys1249}.

\bibitem[{\citenamefont{Barber et~al.}(2018)\citenamefont{Barber, Gibbs, Maeno,
  Mackenzie, and Hicks}}]{barber_resistivity_2018}
\bibinfo{author}{\bibfnamefont{M.}~\bibnamefont{Barber}},
  \bibinfo{author}{\bibfnamefont{A.}~\bibnamefont{Gibbs}},
  \bibinfo{author}{\bibfnamefont{Y.}~\bibnamefont{Maeno}},
  \bibinfo{author}{\bibfnamefont{A.}~\bibnamefont{Mackenzie}},
  \bibnamefont{and} \bibinfo{author}{\bibfnamefont{C.}~\bibnamefont{Hicks}},
  \bibinfo{journal}{Physical Review Letters} \textbf{\bibinfo{volume}{120}}
  (\bibinfo{year}{2018}), ISSN \bibinfo{issn}{0031-9007, 1079-7114}.

\bibitem[{\citenamefont{Maslov and
  Chubukov}(2012)}]{maslov_first-matsubara-frequency_2012}
\bibinfo{author}{\bibfnamefont{D.~L.} \bibnamefont{Maslov}} \bibnamefont{and}
  \bibinfo{author}{\bibfnamefont{A.~V.} \bibnamefont{Chubukov}},
  \bibinfo{journal}{Physical Review B} \textbf{\bibinfo{volume}{86}},
  \bibinfo{pages}{155137} (\bibinfo{year}{2012}), ISSN
  \bibinfo{issn}{1098-0121, 1550-235X},
  \urlprefix\url{https://link.aps.org/doi/10.1103/PhysRevB.86.155137}.

\bibitem[{\citenamefont{Sulewski et~al.}(1988)\citenamefont{Sulewski, Sievers,
  Maple, Torikachvili, Smith, and Fisk}}]{sulewski_far-infrared_1988}
\bibinfo{author}{\bibfnamefont{P.~E.} \bibnamefont{Sulewski}},
  \bibinfo{author}{\bibfnamefont{A.~J.} \bibnamefont{Sievers}},
  \bibinfo{author}{\bibfnamefont{M.~B.} \bibnamefont{Maple}},
  \bibinfo{author}{\bibfnamefont{M.~S.} \bibnamefont{Torikachvili}},
  \bibinfo{author}{\bibfnamefont{J.~L.} \bibnamefont{Smith}}, \bibnamefont{and}
  \bibinfo{author}{\bibfnamefont{Z.}~\bibnamefont{Fisk}},
  \bibinfo{journal}{Physical Review B} \textbf{\bibinfo{volume}{38}},
  \bibinfo{pages}{5338} (\bibinfo{year}{1988}), ISSN \bibinfo{issn}{0163-1829},
  \urlprefix\url{https://link.aps.org/doi/10.1103/PhysRevB.38.5338}.

\bibitem[{\citenamefont{Nagel et~al.}(2012)\citenamefont{Nagel, Uleksin,
  Rõõm, Lobo, Lejay, Homes, Hall, Kinross, Purdy, Munsie
  et~al.}}]{nagel_optical_2012}
\bibinfo{author}{\bibfnamefont{U.}~\bibnamefont{Nagel}},
  \bibinfo{author}{\bibfnamefont{T.}~\bibnamefont{Uleksin}},
  \bibinfo{author}{\bibfnamefont{T.}~\bibnamefont{Rõõm}},
  \bibinfo{author}{\bibfnamefont{R.~P. S.~M.} \bibnamefont{Lobo}},
  \bibinfo{author}{\bibfnamefont{P.}~\bibnamefont{Lejay}},
  \bibinfo{author}{\bibfnamefont{C.~C.} \bibnamefont{Homes}},
  \bibinfo{author}{\bibfnamefont{J.~S.} \bibnamefont{Hall}},
  \bibinfo{author}{\bibfnamefont{A.~W.} \bibnamefont{Kinross}},
  \bibinfo{author}{\bibfnamefont{S.~K.} \bibnamefont{Purdy}},
  \bibinfo{author}{\bibfnamefont{T.}~\bibnamefont{Munsie}},
  \bibnamefont{et~al.}, \bibinfo{journal}{Proceedings of the National Academy
  of Sciences of the United States of America} \textbf{\bibinfo{volume}{109}},
  \bibinfo{pages}{19161} (\bibinfo{year}{2012}), ISSN
  \bibinfo{issn}{0027-8424},
  \urlprefix\url{https://www.ncbi.nlm.nih.gov/pmc/articles/PMC3511099/}.

\bibitem[{\citenamefont{Mirzaei et~al.}(2013)\citenamefont{Mirzaei, Stricker,
  Hancock, Berthod, Georges, Heumen, Chan, Zhao, Li, Greven
  et~al.}}]{mirzaei_spectroscopic_2013}
\bibinfo{author}{\bibfnamefont{S.~I.} \bibnamefont{Mirzaei}},
  \bibinfo{author}{\bibfnamefont{D.}~\bibnamefont{Stricker}},
  \bibinfo{author}{\bibfnamefont{J.~N.} \bibnamefont{Hancock}},
  \bibinfo{author}{\bibfnamefont{C.}~\bibnamefont{Berthod}},
  \bibinfo{author}{\bibfnamefont{A.}~\bibnamefont{Georges}},
  \bibinfo{author}{\bibfnamefont{E.~v.} \bibnamefont{Heumen}},
  \bibinfo{author}{\bibfnamefont{M.~K.} \bibnamefont{Chan}},
  \bibinfo{author}{\bibfnamefont{X.}~\bibnamefont{Zhao}},
  \bibinfo{author}{\bibfnamefont{Y.}~\bibnamefont{Li}},
  \bibinfo{author}{\bibfnamefont{M.}~\bibnamefont{Greven}},
  \bibnamefont{et~al.}, \bibinfo{journal}{Proceedings of the National Academy
  of Sciences} \textbf{\bibinfo{volume}{110}}, \bibinfo{pages}{5774}
  (\bibinfo{year}{2013}), ISSN \bibinfo{issn}{0027-8424, 1091-6490},
  \urlprefix\url{https://www.pnas.org/content/110/15/5774}.

\bibitem[{\citenamefont{Tytarenko et~al.}(2015)\citenamefont{Tytarenko, Huang,
  de~Visser, Johnston, and van Heumen}}]{tytarenko_direct_2015}
\bibinfo{author}{\bibfnamefont{A.}~\bibnamefont{Tytarenko}},
  \bibinfo{author}{\bibfnamefont{Y.}~\bibnamefont{Huang}},
  \bibinfo{author}{\bibfnamefont{A.}~\bibnamefont{de~Visser}},
  \bibinfo{author}{\bibfnamefont{S.}~\bibnamefont{Johnston}}, \bibnamefont{and}
  \bibinfo{author}{\bibfnamefont{E.}~\bibnamefont{van Heumen}},
  \bibinfo{journal}{Scientific Reports} \textbf{\bibinfo{volume}{5}},
  \bibinfo{pages}{12421} (\bibinfo{year}{2015}), ISSN
  \bibinfo{issn}{2045-2322},
  \urlprefix\url{https://www.nature.com/articles/srep12421}.

\bibitem[{\citenamefont{Pustogow et~al.}(2021)\citenamefont{Pustogow, Saito,
  Löhle, Alonso, Kawamoto, Dobrosavljević, Dressel, and
  Fratini}}]{pustogow_rise_2021}
\bibinfo{author}{\bibfnamefont{A.}~\bibnamefont{Pustogow}},
  \bibinfo{author}{\bibfnamefont{Y.}~\bibnamefont{Saito}},
  \bibinfo{author}{\bibfnamefont{A.}~\bibnamefont{Löhle}},
  \bibinfo{author}{\bibfnamefont{M.~S.} \bibnamefont{Alonso}},
  \bibinfo{author}{\bibfnamefont{A.}~\bibnamefont{Kawamoto}},
  \bibinfo{author}{\bibfnamefont{V.}~\bibnamefont{Dobrosavljević}},
  \bibinfo{author}{\bibfnamefont{M.}~\bibnamefont{Dressel}}, \bibnamefont{and}
  \bibinfo{author}{\bibfnamefont{S.}~\bibnamefont{Fratini}},
  \bibinfo{journal}{arXiv:2101.07201 [cond-mat]}  (\bibinfo{year}{2021}),
  \urlprefix\url{http://arxiv.org/abs/2101.07201}.

\bibitem[{\citenamefont{Gingras et~al.}(2019)\citenamefont{Gingras, Nourafkan,
  Tremblay, and Côté}}]{gingras_superconducting_2019}
\bibinfo{author}{\bibfnamefont{O.}~\bibnamefont{Gingras}},
  \bibinfo{author}{\bibfnamefont{R.}~\bibnamefont{Nourafkan}},
  \bibinfo{author}{\bibfnamefont{A.-M.~S.} \bibnamefont{Tremblay}},
  \bibnamefont{and} \bibinfo{author}{\bibfnamefont{M.}~\bibnamefont{Côté}},
  \bibinfo{journal}{Physical Review Letters} \textbf{\bibinfo{volume}{123}},
  \bibinfo{pages}{217005} (\bibinfo{year}{2019}), ISSN
  \bibinfo{issn}{0031-9007, 1079-7114},
  \urlprefix\url{https://link.aps.org/doi/10.1103/PhysRevLett.123.217005}.

\bibitem[{\citenamefont{Suh et~al.}(2020)\citenamefont{Suh, Menke, Brydon,
  Timm, Ramires, and Agterberg}}]{suh_stabilizing_2020}
\bibinfo{author}{\bibfnamefont{H.~G.} \bibnamefont{Suh}},
  \bibinfo{author}{\bibfnamefont{H.}~\bibnamefont{Menke}},
  \bibinfo{author}{\bibfnamefont{P.~M.~R.} \bibnamefont{Brydon}},
  \bibinfo{author}{\bibfnamefont{C.}~\bibnamefont{Timm}},
  \bibinfo{author}{\bibfnamefont{A.}~\bibnamefont{Ramires}}, \bibnamefont{and}
  \bibinfo{author}{\bibfnamefont{D.~F.} \bibnamefont{Agterberg}},
  \bibinfo{journal}{Phys. Rev. Research} \textbf{\bibinfo{volume}{2}},
  \bibinfo{pages}{032023} (\bibinfo{year}{2020}),
  \urlprefix\url{https://link.aps.org/doi/10.1103/PhysRevResearch.2.032023}.

\bibitem[{\citenamefont{Mazin and Singh}(1997)}]{mazin_ferromagnetic_1997}
\bibinfo{author}{\bibfnamefont{I.~I.} \bibnamefont{Mazin}} \bibnamefont{and}
  \bibinfo{author}{\bibfnamefont{D.~J.} \bibnamefont{Singh}},
  \bibinfo{journal}{Physical Review Letters} \textbf{\bibinfo{volume}{79}},
  \bibinfo{pages}{733} (\bibinfo{year}{1997}), ISSN \bibinfo{issn}{0031-9007,
  1079-7114},
  \urlprefix\url{https://link.aps.org/doi/10.1103/PhysRevLett.79.733}.

\bibitem[{\citenamefont{Allen}(2015)}]{allen-2015}
\bibinfo{author}{\bibfnamefont{P.~B.} \bibnamefont{Allen}},
  \bibinfo{journal}{Physical Review B} \textbf{\bibinfo{volume}{92}},
  \bibinfo{pages}{054305} (\bibinfo{year}{2015}).

\bibitem[{\citenamefont{Freericks et~al.}(2001)\citenamefont{Freericks,
  Devereaux, and Bulla}}]{freericks-2001}
\bibinfo{author}{\bibfnamefont{J.~K.} \bibnamefont{Freericks}},
  \bibinfo{author}{\bibfnamefont{T.~P.} \bibnamefont{Devereaux}},
  \bibnamefont{and} \bibinfo{author}{\bibfnamefont{R.}~\bibnamefont{Bulla}},
  \bibinfo{journal}{Physical Review B} \textbf{\bibinfo{volume}{64}},
  \bibinfo{pages}{233114} (\bibinfo{year}{2001}).

\bibitem[{\citenamefont{Li et~al.}(2012)\citenamefont{Li, Gull, and
  Millis}}]{lin-2012}
\bibinfo{author}{\bibfnamefont{N.}~\bibnamefont{Li}},
  \bibinfo{author}{\bibfnamefont{E.}~\bibnamefont{Gull}}, \bibnamefont{and}
  \bibinfo{author}{\bibfnamefont{A.~J.} \bibnamefont{Millis}},
  \bibinfo{journal}{Physical Review Letters} \textbf{\bibinfo{volume}{109}},
  \bibinfo{pages}{106401} (\bibinfo{year}{2012}).

\end{thebibliography}

\end{document}